\documentclass[%
 aip,
 amsmath,amssymb,
 reprint,%
]{revtex4-1}

\usepackage{graphicx}
\usepackage{dcolumn}
\usepackage{bm}

\usepackage[utf8]{inputenc}
\usepackage[T1]{fontenc}
\usepackage{mathptmx}
\usepackage{color}
\usepackage{ulem}
\newcommand{\bra}[1]{\langle\,{#1}\, |}
\newcommand{\ket}[1]{|\,{#1}\,\rangle}
\newcommand{\braket}[2]{\mbox{$\langle\,{#1}\, | \,{#2}\,\rangle$}}

\newcommand{\rvec}[1]{{\mathbf{r}}}

\newcommand{\sub}[2]{{#1}_{\mbox{\!\! \scriptsize #2}}}

\newcommand{\bv}[1]{\mathbf{ #1 }}

\def\beq{\begin{equation}}
\def\eeq{\end{equation}}

\def\CR{\nonumber\\[0.15cm]}

\newcommand{\rref}[1]{Ref.~\cite{#1}}
\newcommand{\fref}[1]{Fig.~\ref{#1}}
\newcommand{\frefp}[2]{Fig.~\ref{#1}~(#2)}

\newcommand{\eref}[1]{Eq.~(\ref{#1})}

\newcommand{\sref}[1]{section~\ref{#1}}

\newcommand{\cref}[1]{chapter~\ref{#1}}

\newcommand{\Cref}[1]{Chapter~\ref{#1}}
\newcommand{\tref}[1]{table~\ref{#1}}

\newcommand{\aref}[1]{appendix~\ref{#1}}
\newcommand{\bref}[1]{(\ref{#1})}

\usepackage{ulem}  
\begin{document}

%
\preprint{AIP/123-QED}

\title[Excitation Transport in Molecular Aggregates with thermal motion]{Excitation Transport in Molecular Aggregates with thermal motion}

\author{R. Pant}
 \email{ritesh17@iiserb.ac.in}
\affiliation{Department of Physics, Indian Institute of Science Education and Research, Bhopal, Madhya Pradesh 462 023, India}
\author{S.~W\"uster}
 \email{sebastian@iiserb.ac.in}
\affiliation{Department of Physics, Indian Institute of Science Education and Research, Bhopal, Madhya Pradesh 462 023, India}

\date{\today}

\begin{abstract}
Molecular aggregates can under certain conditions transport electronic excitation energy over large distances due to dipole-dipole interactions.
Here, we explore to what extent thermal motion of entire monomers can guide or enhance this excitation transport. The motion induces changes of aggregate geometry and hence modifies exciton states. Under certain conditions, excitation energy can thus be transported by the aggregate adiabatically, following a certain exciton eigenstate.
While such transport is always slower than direct migration through dipole-dipole interactions, we show that transport through motion can yield higher transport efficiencies in the presence of on-site energy disorder than the static counterpart.
For this we consider two simple models of molecular motion: (i) longitudinal vibrations of the monomers along the aggregation direction within their inter-molecular binding potential and (ii) torsional motion of planar monomers in a plane orthogonal to the aggregation direction. The parameters and potential shapes used are relevant to dye-molecule aggregates. We employ a quantum-classical method, in which molecules move through simplified classical molecular dynamics, while the excitation transport is treated quantum mechanically using Schr\"odinger's equation. For both models we find parameter regimes in which the motion enhances excitation transport, however these are more realistic for the torsional scenario, due to the limited motional range in a typical Morse type inter-molecular potential. We finally show that the transport enhancement can be linked to adiabatic quantum dynamics. This transport enhancement through adiabatic motion appears a useful resource to combat exciton trapping by disorder.
\end{abstract}

\maketitle
\section{Introduction} 
%
Molecular aggregates in which a large number of organic molecules assemble into a fairly regular structure can exhibit significant excitation energy transport along the structure \cite{brixner2017exciton, haedler2016pathway}, which plays a key role in photosynthetic light harvesting processes \cite{grondelle:book,grondelle:review} and has the potential for technological exploitation, e.g.~in dye-sensitized solar cells \cite{zhang2017dye, ghosh1978merocyanine} or thin-film optical and optoelectronic devices \cite{malyshev2000intrinsic}. In all of these, molecular aggregates facilitate the absorption of light and subsequent transfer of the absorbed energy to a reaction centre \cite{saikin2013photonics, macedo2019perylene} in the form of an electron-hole pair known as exciton. This transfer of excitation relies on the long range dipole-dipole interactions between the monomers in the aggregate. 

Such dipole-dipole interactions are also a characteristic feature of Rydberg aggregates \cite{wuster2018rydberg}, which hence have been proposed as quantum simulators for molecular aggregates \cite{schonleber2015quantum,hague2012quantum}. In Rydberg aggregates, a chain of highly excited Rydberg atoms transports a single energy quantum on spatial- and temporal scales quite different from the molecular context. While the excitation transfer process in molecular aggregates is typically strongly affected by decoherence \cite{kasha1963energy, chen2011excitation,roden2009influence,haken1972coupled}, it barely is in ultra-cold atoms, as has been experimentally demonstrated \cite{barredo:trimeragg,labuhn:rydberg:ising,Marcuzzi:manybodyloc}.

\begin{figure}[htb]
\includegraphics[width=0.99\columnwidth]{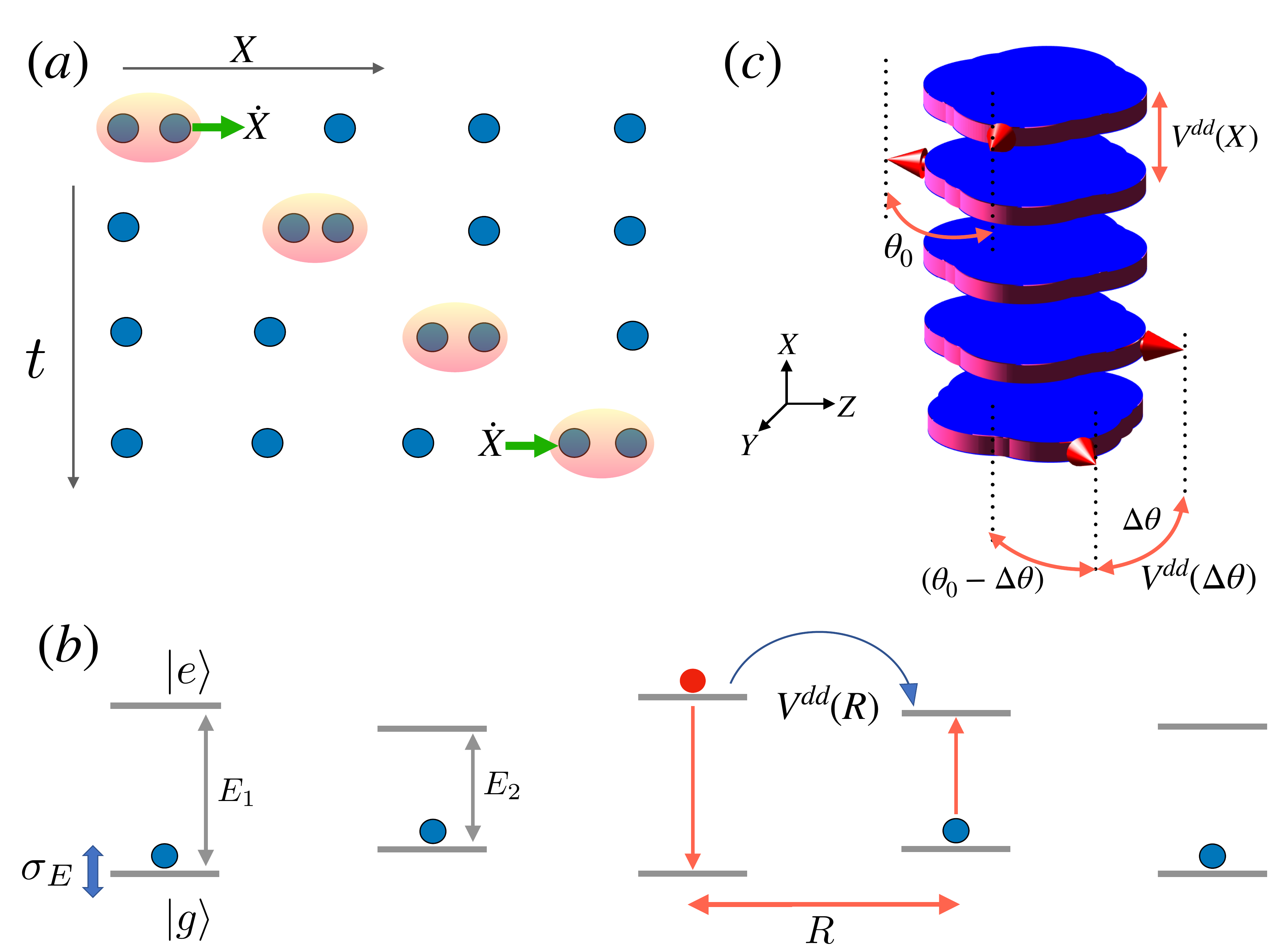}
\caption{\label{overfig} (a) Interplay of molecular motion and excitation transfer in adiabatic excitation transport. Blue $\bullet$ are molecules or atoms, $X$ indicates their position and $\dot{X}$ (green arrow) their velocity. The orange shade represents a selected de-localized exciton state. This state is always
located on the two \emph{closest} molecules, such that if the state is adiabatically followed, the indicated motion entails excitation transport.
 (b) Energy level schematic, with molecular electronic ground state $\ket{g}$ and excited state $\ket{e}$ and dipole-dipole interactions $V^{dd}(R)$. We also sketch on-site disorder $\sigma_E$ of transition energies. (c) More detailed sketch of a 1D chain of planar molecules with torsional motion along $\theta$ and longitudinal motion along $X$. (blue disk) molecules, (red large arrow) direction of the transition dipole moment axis between $\ket{g}$ and $\ket{e}$. $\theta_0$ is the relative nearest neighbour angle in equilibrium and $\Delta \theta$ possible angular disorder around this value.
}
\end{figure}

An idea that naturally arises in Rydberg aggregates, is adiabatic excitation transport through atomic motion \cite{wuster2010newton,wuster2018rydberg, mobius2011adiabatic}. In adiabatic excitation transport, slow motion of the atoms combined with excitation transport via dipole-dipole interactions can result in efficient and guided transport of the excitation from one end of an atomic chain to the other, see schematic in \frefp{overfig}{a}. Based on the analogy between Rydberg- and Molecular aggregates, the question then arises whether adiabatic excitation transport can play a functional role in molecular aggregates, e.g.~for light harvesting.
Here we report initial explorations of this idea, ignoring for now the effect of the \emph{intra-molecular} vibrations but including some effects of the protein environment as a static disorder in energy. Our model is then a closed quantum system for the excitation transport, and our main observables are derived from the dynamics of excitation population on different molecules, classically averaged over the disordered ensemble. The motivation for this framework is as follows: quantum adiabatic following in excitation transport was first reported in the atomic case where internal vibrations are absent \cite{wuster2010newton}. If we cannot find similar features in a molecular setting excluding internal vibrations, they are unlikely to be present in a case where vibrations are included, since these are known to significantly modify excitation transport \cite{chen2011excitation, roden2009electronic, roden2012accounting}. We however shall find that adiabatic excitation transport persists and thus intend to explore in the future to what extent adiabatic excitation transport can survive the coupling to internal molecular vibrations. Meanwhile, our results should already be applicable to some extent to those molecular aggregates where the coupling of excitons with internal vibrations is weak, such as those reported in \cite{eisfeld2002j, haedler2015long,kang2019ultrafast}.
 
Three key features, which are  the focus of the present article, change the physics of transport in molecular aggregates compared to the simpler ultra-cold atomic scenario even if intra-molecular vibrations are neglected. These features are site-to-site energy disorder, inter-molecular binding and random thermal positions and velocities. Varying all parameters pertaining to these within ranges relevant for molecular aggregates, we map out regimes where excitation transport involving molecular motion can yield higher transport efficiencies than direct dipole-dipole transport in the immobile case, since motion counters energy disorder.

For this we set up two different simple models for molecular motion in aggregates: (i) Longitudinal motion, in which molecules move classically along the direction of aggregation only, bound to their neighbors through a Morse potential. This motion affects the dipole-dipole Hamiltonian through varying distances between molecules. (ii) Torsional motion, in which molecules at fixed separation can rotate in the plane orthogonal to the aggregation direction, which affects the dipole-dipole Hamiltonian through varying angles between transition dipole moments. Both models also involve a quantum degree of freedom for the electronic state that allows for a single, possibly delocalized electronic excitation. We find that excitation transport is more positively affected by motion in the torsional model, 
since for a realistic motional range of molecules larger variations of dipole-dipole interactions and hence exciton states are accessible through varying angles between dipole moments. In comparison, during longitudinal motion, variations due to changing separations between monomers are smaller.
Using a measure for the adiabaticity of quantum transport that we propose and benchmark in \cite{pant:adiabaticity}, we finally show that the increase of transport efficiency due to motion can at least partially be attributed to adiabatic quantum dynamics.

The effect of molecular motion on excitation transport has also been investigated in \cite{Asadian_2010,semiao2010vibration,behzadi2017effects,o2014non,mulken2011directed}. Most of these studies consider transport in the presence of decoherence and none explore the aspect of adiabaticity, as we do here. In contrast to our explicit model for thermal molecular motion, Refs.~\cite{Asadian_2010,behzadi2017effects} constrain classical harmonic motion to selected harmonic normal modes. In \rref{semiao2010vibration,o2014non} inter-molecular vibrations are considered quantum mechanically, focussing mainly on one relevant resonant mode. All articles find an increased transport efficiency in certain parameter regimes when comparing a static with a mobile scenario, in agreement with the results that we shall present. The influence of adiabatic conformational change on exciton migration in semiconducting polymers has been investigated in Refs.~\cite{binder2018conformational,binder2019first}, which convey a similar picture as found here, including deviations from adiabaticity. Similar studies for electron transfer between an organic sensitizer molecule and a semiconductor surface show that both adiabatic and non-adiabatic pathways are available for electron transfer and both are significant \cite{stier2002nonadiabatic, duncan2005nonadiabatic, duncan2005ab}. Another context where adiabaticity could form a tool for transport enhancement is additional laser driving of light harvesting molecules \cite{Dijkstra_transport_STIRAP}.

This article is organised as follows: In \sref{Models and Methods}, we introduce the features of our molecular aggregate model that are common to both scenarios listed above (longitudinal and torsional motion), such as dipole-dipole interactions, mechanical motion and the quantum-classical propagation scheme that we employ. The remainder of the article is then arranged in two parts, in \sref{Long_Cradle} we explore motion of monomers along the aggregation axis, while in \sref{Rot_Cradle} monomers rotate in a plane orthogonal to that axis. Both sections are then structured similarly: We firstly demonstrate in one clear but not necessarily realistic scenario how adiabatic excitation transport would proceed in a molecular setting (\sref{AET_long} and \sref{AET_rot}), followed by an extensive parameter survey comparing the transport efficiency with and without motion (\sref{Transport_Efficiency} and \sref{RT}). In a final subsection for each part we analyze in detail to which extent the transport can be traced back to adiabatic changes of the exciton Hamiltonian (\sref{Transport_adiabaticity_long} and \sref{Transport_adiabaticity_rot}). A crucial feature in our survey is on-site disorder, which we introduce in \sref{Localization} and then use also in \sref{Rot_Cradle}. Finally the appendices contain details on our estimates of moments of inertia, \aref{MOI_rot}, single trajectory simulations for the case of longitudinal motion, \aref{Long_single_traj},  and torsional motion, \aref{rot_single_traj}, as well as measures of adiabaticity, \aref{allowed_jumps}.

\section{Excitation transport and molecular motion} \label{Models and Methods}
%
We model $N$ monomers with mass $M$ and moment of inertia $I$, arranged in a one dimensional (1D) chain along the $X$ direction, where the $n$'th monomer is located at a definite, classical position $X_n$. These monomers can be bound to each other by van-der-Waals forces and/or hydrogen bonds, with inter monomer distances of the order of Angstr\"om \cite{may2011charge}. We consider each monomer as an electronic two level system with ground state $\ket{g}$ and first excited state $\ket{e}$. The transition dipole moment $\boldsymbol{\mu}$ between these two states is assumed fixed in the $YZ$-plane orthogonal to the aggregation direction, and at an angle $\theta$, wrt.~the $Z$ axis, see \frefp{overfig}{c}. The distance between monomers shall be large enough to neglect electronic wave function overlap, so that there is no direct exchange of electrons between the monomers \cite{may2011charge}. Therefore the only interactions capable of excitation energy transfer are Coulomb interactions. For the large distances between the monomers, we assume that these can be approximated by the dipole-dipole interaction Hamiltonian $\sub V{mn}^{(dd)}(\mathbf{R})$, between monomers $m$ and $n$, which reads
\begin{eqnarray}\label{dip-dip}
\sub V{mn}^{(dd)}(\mathbf{R}) = \frac{{\pmb{\mu}}_m \cdot \pmb{\mu}_n}{|{\mathbf{X}_{mn}}|^3} - 3 \frac{(\pmb{\mu}_m \cdot {\bv{X}}_{mn}) (\pmb{\mu}_n \cdot {\bv{X}}_{mn})}{ |{\bv{X}_{mn}}|^5 },
\end{eqnarray}
where $\mathbf{R} = (X_1, X_2, ..., X_N, \theta_1, \theta_2, ..., \theta_N)^T$ denotes the collection of all molecular coordinates, including locations, $X_n$, and angles, $\theta_n$, of dipole moments wrt.~to an axis orthogonal to the aggregation direction, see \frefp{overfig}{c}. Further, $\mathbf{X}_{mn}=\mathbf{X}_m-\mathbf{X}_n$ is the separation of monomer $m$ and monomer $n$ and ${\pmb{\mu}}_m=\pmb{\mu}_m(\theta_m) $ and $\pmb{\mu}_n=\pmb{\mu}_n(\theta_n) $ are their transition dipole moments. We shall assume that there is only a single excitation present in the aggregate, hence the electronic Hilbert space is spanned by $\ket{m} = \ket{g...e...g}$, where only the $m$'th molecule is in the excited state and all others are in their ground state. We call $\{\ket{m}\}$ the diabatic basis. While we included longitudinal coordinates $X_n$ and torsional coordinates $\theta_n$ simultaneously in \bref{dip-dip}, we shall only present their dynamics one-by-one in this article, in order to separately assess their potential for enhancing transport, and comment on joint dynamics of both degrees of freedom in the conclusion.

With the above restrictions on degrees of freedom, the Hamiltonian of our system can be written as \cite{ates2008motion}
\begin{eqnarray}\label{Hamiltonian}
H(\mathbf{R}) = - \sum_{n=1}^{N}\left[ \frac{\hbar^2}{2M} \frac{\partial^2}{\partial X_n^2} + \frac{\hbar^2}{2I} \frac{\partial^2}{\partial \theta_n^2} \right] + H_{el}(\mathbf{R}),
\end{eqnarray}
where the first term gives the kinetic and rotational energy of the molecules and $H_{el}(\mathbf{R})$ is single exciton Hamiltonian, given by \cite{kuhn2011quantum}
\begin{align}
\label{single_exciton_Hamiltonian}
H_{el}(\mathbf{R}) &= \sum_{m=1}^{N}E_{m} \ket{m}\bra{m} +  \sum_{nm;n\neq m}V_{mn}^{(dd)}(\mathbf{R}) \ket{m}\bra{n}\CR 
&+ \left( \frac{1}{2}\sum_{nm;n\neq m} V_{mn}^{(B)}(\mathbf{R})  \right)\mathbb{I}.
\end{align}
Here $E_{m}$ is the electronic excitation energy at site m, and $V_{mn}^{(dd)}(\mathbf{R})$ is the matrix element for the dipole-dipole interaction between monomer $m$ and $n$ given in \eref{dip-dip}, which is responsible for excitation energy transfer. The transition energy $E_{m}$ at each site is different since the influence of the environment could be different for each monomer, see e.g.~\rref{wang2015open,AdRe06}. $V_{mn}^{(B)}(\mathbf{R})$ denotes interactions that do not depend on the electronic state, which we assume to be the case for the inter-molecular binding potential, for simplicity.

To study the dynamics induced by the Hamiltonian \bref{Hamiltonian}, consider the eigenstate of the electronic part $H_{el}(\mathbf{R})$
\begin{eqnarray}\label{adiab_evalue_equation}
H_{el}(\mathbf{R}) \ket{\varphi_s{(\mathbf{R})}} = U_s(\textbf{R}) \ket{\varphi_s{(\mathbf{R})}},
\end{eqnarray}
where $U_s(\mathbf{R})$ defines the $s$'th potential energy surface for a given molecular configuration. $U_s(\mathbf{R})$ is called the adiabatic potential energy surface and the corresponding eigenstates are referred as adiabatic basis states or Frenkel excitons. Note that $U_s(\mathbf{R})$ are \emph{supra-molecular energy surfaces}. Each adiabatic state can be written in terms of the diabatic states as, 
\begin{eqnarray}\label{adiab_state}
\ket{\varphi_s{(\mathbf{R})}} =  \sum_{m=1}^{N} f_{sm} (\mathbf{R}) \ket{m}.
\end{eqnarray}
We use a mixed quantum-classical method, Tully's surface hopping \cite{tully1990molecular,tully1971trajectory}, where the motion of molecules is treated classically according to Newton's equations
\begin{eqnarray}
\label{newton_longit}
M \frac{\partial^2}{\partial t^2}{X}_m = - \nabla_{X_m} U_s(\textbf{X}) - \sum_n \nabla_{X_m} V_{mn}^{(B)} (\mathbf{X}),
\end{eqnarray}
and an ensemble of trajectories is propagated on a specific Born-Oppenheimer surface $U_s$. The surface index $s$ is allowed to stochastically jump in time, 
to take into account non-adiabatic transitions from one surface to another and the corresponding change of forces acting on the molecules. Here, we show \bref{newton_longit} for the case of longitudinal motion $\frac{\partial^2 {X}_m}{\partial t^2}$ only, its torsional version shall be given and used in \sref{Rot_Cradle}.

Expanding the total wavefunction of the system in the adiabatic basis defined above $\ket{\psi(\mathbf{X},t)} = \sum_{m=1}^{N} \tilde{c}_m \ket{\varphi_m{(\mathbf{X})}}$, we can also obtain the following Schr\"odinger equation for the complex amplitudes $\tilde{c}_m$,
\begin{eqnarray}
\label{TDSE_adiabatic_basis}
i \frac{\partial}{\partial t} \tilde{c}_m = U_m(\textbf{X}) \tilde{c}_m - i \sum_{n=1}^{N} d_{mn} \tilde{c}_n,
\end{eqnarray}
where $d_{mn}$ are the non-adiabatic coupling coefficients, which also control the probability of stochastic jumps between surfaces in Tully's algorithm.
They can be written as 
\begin{eqnarray}
\label{d_mn}
d_{mn} = - \frac{1}{M} \bra{\varphi_m(\textbf{X}(t))} \pmb{\nabla}_{\mathbf{X}} \ket{\varphi_n(\textbf{X}(t))} \cdot \frac{\partial {X}_m}{\partial t}.
\end{eqnarray}
Besides the movement of the molecules, we are interested in the exciton dynamics, for which we evolve the total wave function in the diabatic basis $\ket{\psi(\mathbf{X},t)} = \sum_{m=1}^{N} c_m(t) \ket{m}$, instead of \eref{TDSE_adiabatic_basis}. Its time evolution is thus determined by
\begin{eqnarray}
\label{TDSE_diabatic_basis}
i \frac{\partial}{\partial t} {c}_m = \sum_{n=1}^{N} H_{mn}[X_{mn}(t)] {c}_n, 
\end{eqnarray}
and coupled to \bref{newton_longit} through the dependence of the electronic Hamiltonian on all molecular positions. Here $H_{mn}[X_{mn}(t)]$ is the matrix element for the electronic coupling in \eref{single_exciton_Hamiltonian}, with the dipole-dipole interaction given by \eref{dip-dip_long} and $X_{mn}=|\mathbf{X}_m-\mathbf{X}_n|$. In \bref{TDSE_diabatic_basis} we thus evolve the electronic state in the diabatic basis, which is precisely defined in our case and an equally efficient choice for solving the TDSE as the adiabatic basis, in contrast to some quantum chemistry problems \cite{xie2018performance, liang2018symmetrical}. It then offers conceptual advantages in avoiding geometric phase issues.

While we outlined the formalism jointly here for molecular degrees of freedom $X_n$ and $\theta_n$, the rest of our study is arranged in two parts where we first 
assume a fixed direction of the molecular transition dipoles, but allow monomers to move along the aggregation direction, and in a second part fix molecular positions along the aggregate axis, but allow their torsional motion through a plane orthogonal to that axis, and hence varying transition dipole directions.
This splitting has the objective to clearly determine which degrees of freedom are more conducive for motional enhancement of excitation transport.

\section{Excitation transport by longitudinal motion}
\label{Long_Cradle}
\begin{figure}[htb]
\includegraphics[width=0.99\columnwidth]{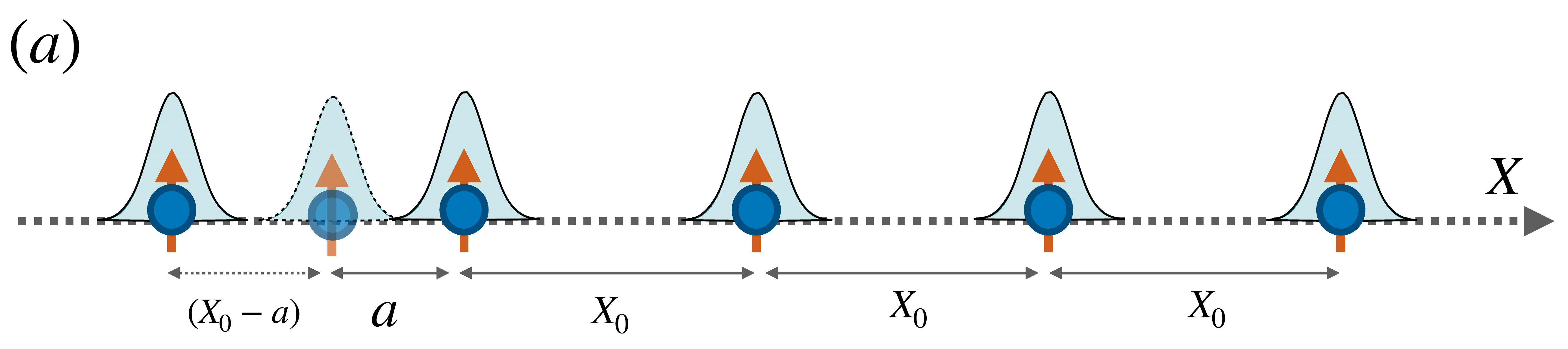}
\includegraphics[width=0.99\columnwidth]{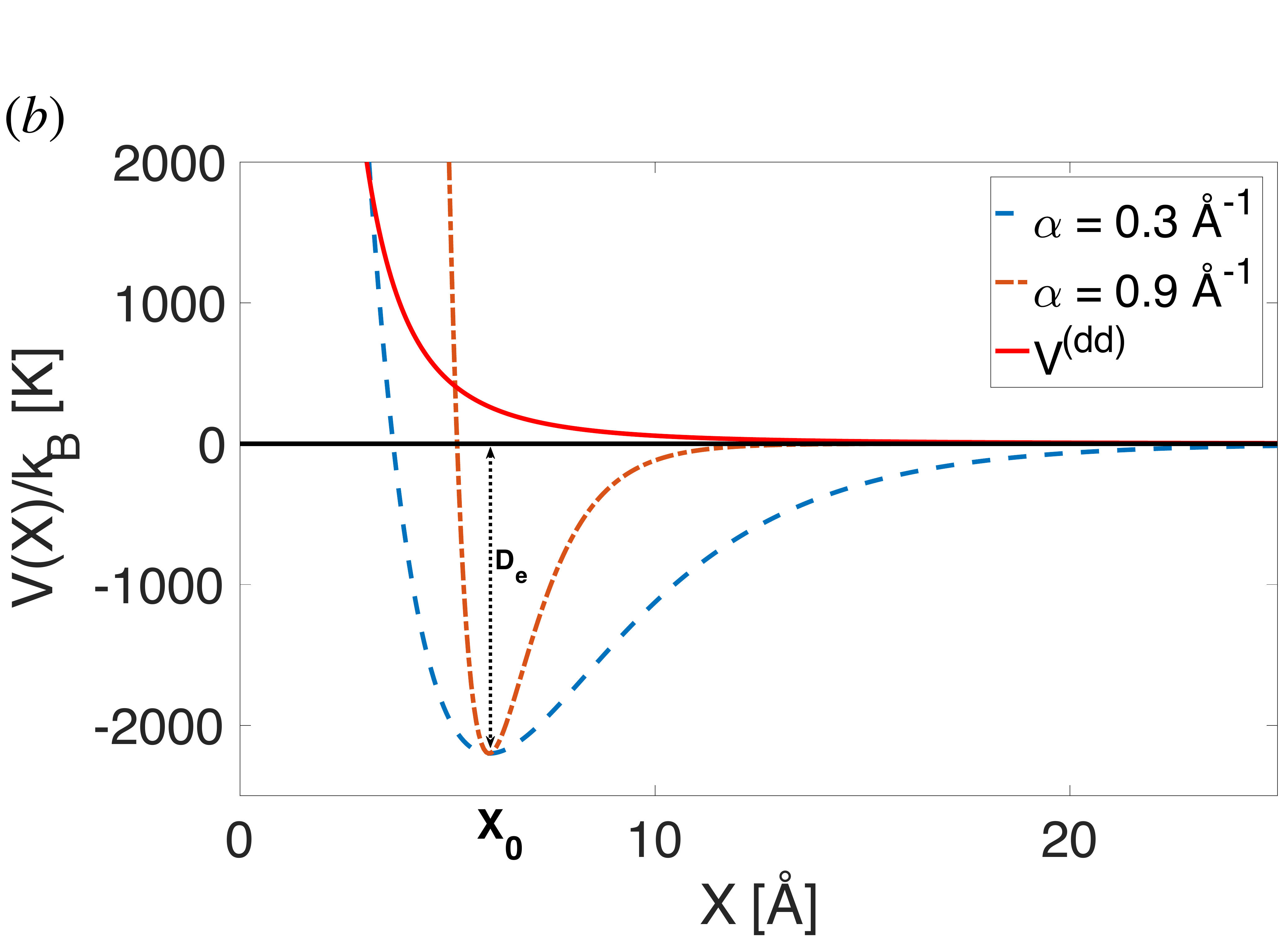}
\caption{\label{Long_oneD} (a) Sketch of 1D molecular aggregate with off diagonal disorder, all transition dipoles are parallel. Shaded Gaussians indicate possible initial thermal position disorder of each monomer. (b) Inter-molecular potentials: Morse potential for $\alpha=0.3$ \AA$^{-1}$ (blue dashed) and $\alpha=0.9$ \AA$^{-1}$ (red dot-dashed) and the strength of dipole-dipole interactions $V_{nm}^{(dd)}$ (red solid line).}
\end{figure}

%
In this section, we consider a chain of identical monomers with all transition dipole moment axes $\pmb{\mu}$ fixed orthogonal to the chain axis. This is commonly referred to as H-aggregate \cite{rosch2006fluorescent}, and might be more suitable for excitation transport than J-aggregates with head to tail $\pmb{\mu}$, due to the larger excited state lifetime \cite{kasha1965exciton}. The monomers can move along the $X$ axis joining them, see \frefp{Long_oneD}{a}. Dipole-dipole interactions \eref{dip-dip} in this case are
\begin{eqnarray}\label{dip-dip_long}
\sub V{mn}^{(dd)}(\mathbf{X}) = \frac{\mu^2}{|{\mathbf{X}_{mn}}|^3}.
\end{eqnarray}
The monomers in aggregates are bound by van-der-Waals forces, which we model with a Morse potential 
\begin{eqnarray}
\label{Morse_potential}
\sub V{mn}^{(B)}(\mathbf{X}) = D_e\Big[ e^{-2\alpha(|\mathbf{X}_{mn}| - X_0)} - 2 e^{-\alpha(|\mathbf{X}_{mn}| - X_0)} \Big],
\end{eqnarray}
where $D_e$ is the depth of the well, $X_0$ the equilibrium distance and $\alpha$ controls the width of the potential. The smaller $\alpha$, the softer and wider is the potential, see \frefp{Long_oneD}{b}.
To be specific, we choose a mass $M = 902330$ a.u. and a transition dipole moment $\mu = 1.12$ a.u., such that nearest neigbor dipole-dipole interactions roughly match the real ones in e.g.~carbonyl-bridged triaryl-amine (CBT) dyes \cite{saikin2017long}.

\subsection{Idealized adiabatic excitation transport} \label{AET_long}

In this section, we first elucidate the concept of adiabatic excitation transport in a clear cut, albeit constructed case.
For this, all molecules are initially placed at the equilibrium separation ($X_0= 6$\AA) of the Morse potential, except the first two molecules, which have a closer separation $a=4$\AA, see \frefp{Long_oneD}{a}. For this configuration the dipole-dipole interaction between the first two molecules is much stronger than between any other neighboring molecules in the chain. This results in the localization of the excitation on these first two molecules, such that the initial electronic aggregate state is to a very good approximation the exciton
\begin{eqnarray}\label{adiab_estate}
\ket{\psi(0)} = \frac{1}{\sqrt{2}} (\ket{m=1} + \ket{m=2}).
\end{eqnarray}
A second important consequence of the initial condition, is that the first two molecules very strongly repel each other, since they are deep on the inner repulsive side of the Morse potential in 
 \frefp{Long_oneD}{b}. Around the geometry choice described so far, the initial positions and velocities of the molecules are randomized according to the Maxwell Boltzmann distribution \cite{footnote:MBdistribution} at temperature $T=300$ K. However the standard deviation in position due to thermal motion is $0.5$ \AA, which is fairly very small compared to the dislocation imposed on the first two sites i.e., 4 \AA. Finally, we neglect on-site disorder in this section, such that $E_m = 0$ in \eref{single_exciton_Hamiltonian}.

\begin{figure}[htb]
\includegraphics[width=0.99\columnwidth]{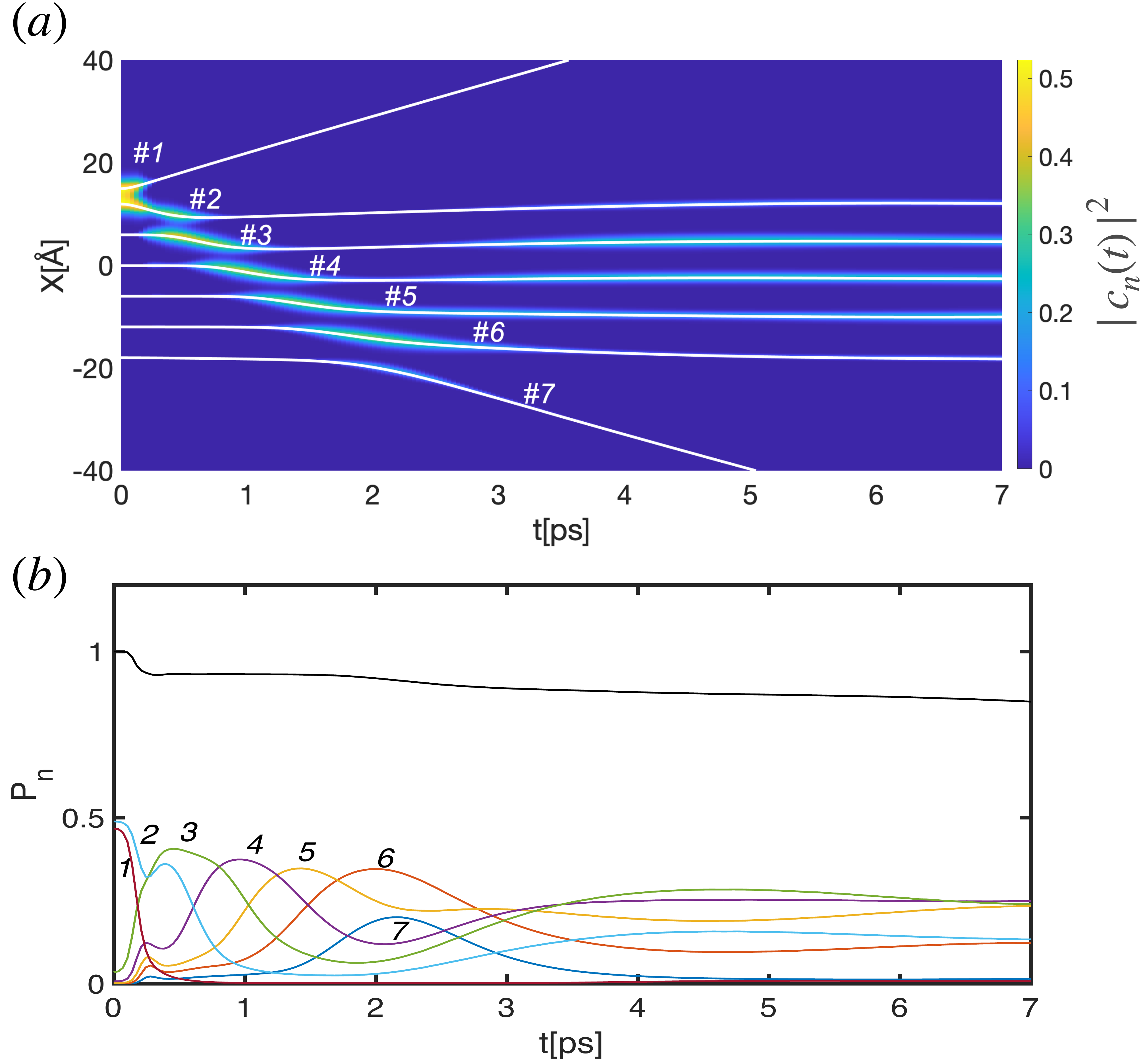}
\caption{\label{adiabatic_long} Exemplary adiabatic excitation transport with molecules. (a) (solid white) Mean trajectories of individual molecules, with site index $n$ as indicated. (color shading) Mean electronic excitation probabilities or diabatic populations $|c_n|^2$ are indicated by color around the respective trajectory. (b) (colored numbered lines)  Individual diabatic populations $|c_n|^2$, (black solid) adiabatic population on the initial surface. Binding parameters are $\alpha = 0.5$ \AA$^{-1}$ and a well depth of $D_e=2200$ K. Results are averaged over $10^5$ trajectories.}
\end{figure}

The resultant motion of the molecules and the excitation transfer are shown in \fref{adiabatic_long}, using \eref{newton_longit} coupled to \eref{TDSE_diabatic_basis}.
Initially the two closest molecules strongly repel each other. Molecule 2 moves towards molecule 3, while molecule 1 escapes  the chain, since
the initial potential energy $\sub V{12}^{(B)}(a)$ by far exceeds the binding energy $D_e$. When molecule 2 reaches molecule 3 those two constitute the new closest proximity pair. 
Since the motional time-scale $\sub{\tau}{mov}$ is large compared to the characteristic time-scale for dipole-dipole interactions $\tau_{dd}\equiv \pi/\sub V{12}^{(dd)}(a)=2.7$ fs, the system can adiabatically follow the exciton quantum state that initially corresponds to \eref{adiab_estate}, which is always localized on the two closest molecules. Around $t=0.5$ ps, it is hence now localized on molecule 2 and 3. Since in this close encounter, also the momentum is transferred from molecule 2 to molecule 3, the process continues along the chain until molecule 7 escapes it in the end. Just prior to that, at $t=2 ps$ in \fref{adiabatic_long}, the excitation has to a large extent been transported to the end of the chain, on molecules 6 and 7.
We see a small change in the adiabatic population, black line in \frefp{adiabatic_long}{b}, which reduces the transport fidelity. Otherwise, our extreme choice of initial conditions has replicated the near perfect transport scenario of the atomic case \cite{wuster2010newton}, where cold atoms are not bound to their neighbor. A very similar scenario arises if we start from the initially localized state $\ket{\psi(0)}=\ket{1}$. Since this can be written as a linear combination of \bref{adiab_estate} and the corresponding anti-symmetric exciton, both of which are adiabatically transported to the end of the chain as in \fref{adiabatic_long}, also the excitation from initial state $\ket{1}$ reaches the end of the chain through the motion.

In summary, a pulse combining motion and excitation transfer can facilitate high fidelity transport of an excitation through a chain. For its kinematic similarity with the popular class-room tool to demonstrate momentum conservation, the process has been likened to Newton's cradle in \rref{wuster2010newton}. The physical basis is quantum adiabaticity, which leads to a limitation of this technique: To remain adiabatic, we require slow motion  $\sub{\tau}{mov}\gg\tau_{dd}\equiv \pi/\sub V{12}^{(dd)}(a)$, as discussed above, which means the transported excitation energy will always arrive earlier if we start in a localized state $\ket{\psi(0)}=\ket{\pi_1}$ instead of \eref{adiab_estate}, and consider an equidistant chain. However, the situation is less clear when on-site energy disorder is present, since localization might then preclude an excitation starting in a localized state to arrive at all. 

The example in this section has been chosen to clearly illustrate the concept of adiabatic excitation transport, but will not likely be practically useful due to the extreme and thus thermally inaccessible initial state for the positions of monomers 1 and 2. In the following we much more generally compare the transport properties of moving and static aggregates in the presence of on-site energy disorder.

\subsection{Disorder and exciton localization} \label{Localization}

The on-site energy disorder $\sigma_E$ in an aggregate, sketched in \frefp{overfig}{b}, arises due to the coupling of the monomers with their environment~\cite{wang2015open,AdRe06}. 
Since the local environment may be different for each site, this can cause slightly different transition energy shifts $E_n$ for each monomer. We assume here that the time-scale for variation of such shifts is slow, so that the $E_n$ are constant throughout the transport process, hence we only treat static disorder. For the $E_n$ in \eref{single_exciton_Hamiltonian}, 
we assume a Gaussian distribution \cite{valleau2012exciton}
\begin{eqnarray}\label{Energy_Distribution}
p_E(E_n - E_0) = \frac{1}{\sqrt{2\pi \sigma_E}} e^{-(E_n - E_0)^2/2\sigma_E}
\end{eqnarray}
where $\sigma_E$ is the standard deviation and $E_0$ the unperturbed transition energy of each molecule. The distribution is assumed to be identical for all monomers. For realistic systems, more sophisticated distributions may apply \cite{eisfeld2010excitons, hestand2018expanded}.

Disorder strongly affects the energy level structure and wave-function of exciton states in \bref{adiab_evalue_equation}, which in turn influences the excitation transfer. One measure of the impact of disorder is the de-localization length of the exciton over the aggregate \cite{meier1997polarons, ray1999short}. For weak disorder the exciton is de-localized over the entire aggregate, while for strong disorder it becomes localized on a smaller number of monomers. This may cause exciton trapping \cite{fidder1991dynamics, fidder1991optical}, which is detrimental to excitation transport.

We now demonstrate that motion can help excitation transport overcome disorder induced localization in a simple test-case. For this, we take a chain composed of seven monomers placed at their equilibrium separations $X_0$, that are subject to thermal position and velocity distributions. We compare a static and a mobile scenario, both exhibiting an identical realisation of the disorder \bref{Energy_Distribution}.

\begin{figure}[htb]
\includegraphics[width=0.99\columnwidth]{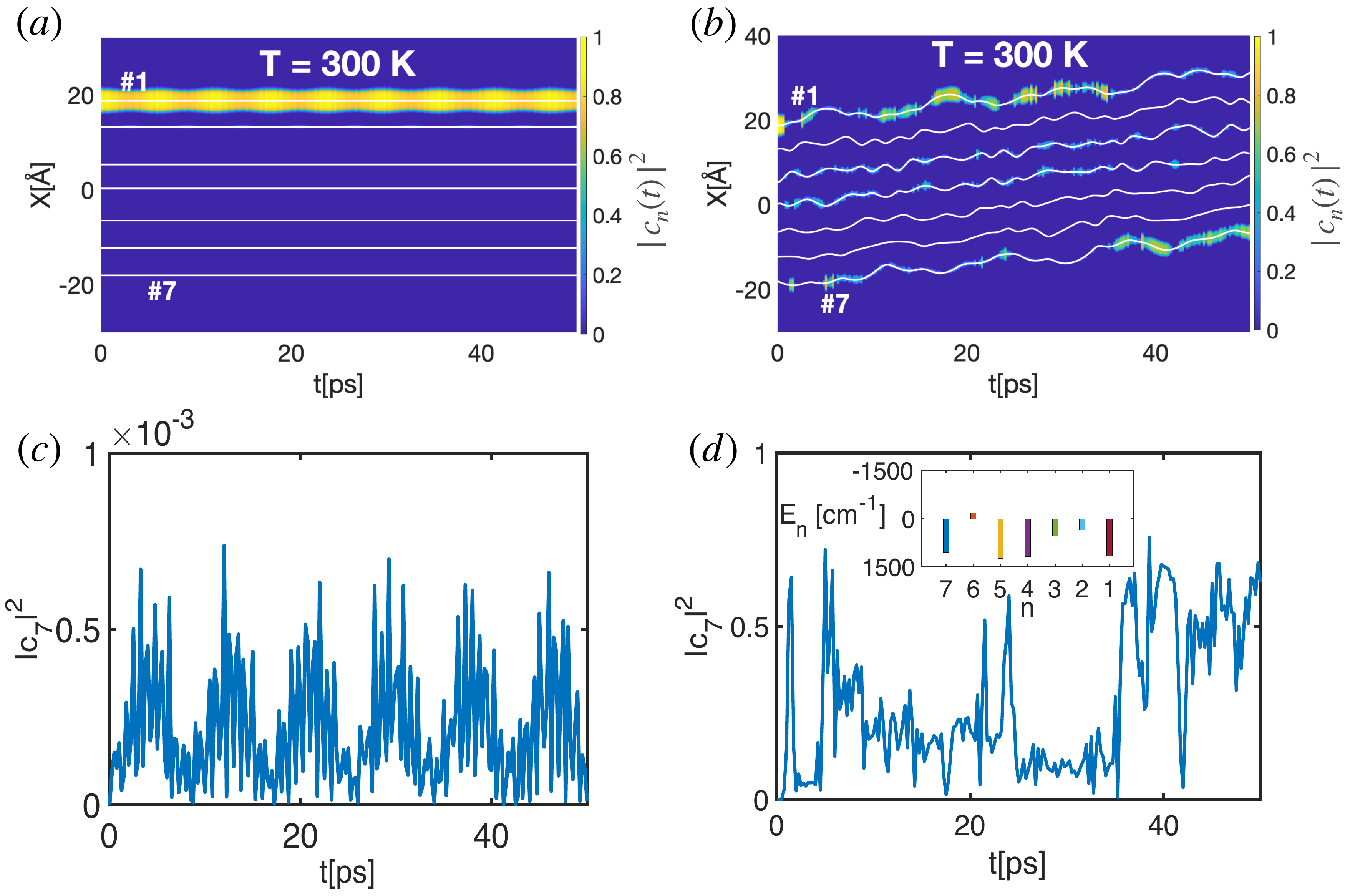}
\caption{\label{ST_longit} Excitation transport can overcome on-site energy disorder through motion. (a) Monomer trajectories and excitation probability in the immobile case in the same style as for \fref{adiabatic_long}. (b) The same for mobile molecules. (c) Population reaching the output site \#7 in the immobile case. (d) The same for the mobile case. The inset shows the realisation of on-site energy disorder \bref{Energy_Distribution} used for both, panel (a) and (b).}
\end{figure}

The dynamics of excitation transfer for a single trajectory without any initial close proximity of the first two molecules ($a=X_0$) is shown in \fref{ST_longit}. The disorder strength was $\sigma_E = 500$ cm$^{-1}$, $\alpha = 0.5$\AA$^{-1}$  and the temperature $T = 300$ K. At $t=0$ the excitation is injected at the first site (\#1, input site), hence
\begin{eqnarray}\label{single_site_inistate}
\ket{\psi(0)} =\ket{m=1}.
\end{eqnarray}
in contrast to \sref{AET_long}. We choose \bref{single_site_inistate} for simplicity. Starting from this initial state, we investigate whether the excitation reaches the output site. We see that for the static system with $\dot{\mathbf{X}}=0$, the population reaching the output site remains very small \fref{ST_longit}(c). In contrast, for the dynamic system thermal fluctuations overcome the disorder and can deliver the excitation to the output site with quite high probability.
While here we only demonstrated this for a specific single realisation of random positions, velocities and energy disorder, the latter shown in \frefp{ST_longit}{d}, we will nextly confirm that motion can help to overcome disorder also in the ensemble average. A detailed inspection of the adiabatic populations $\tilde{p}_n=|\tilde{c}_n|^2$, see \bref{TDSE_adiabatic_basis}, shows however that the electronic dynamics is only intermittently adiabatic, with a large number of non-adiabatic transitions. We will comment on this again later in this article.

\subsection{Transport efficiency} \label{Transport_Efficiency}

The single trajectory simulation in the previous section indicates that the motion of the molecules in an aggregate can potentially play a key role in transport of excitation energy in the presence of disorder. To explore this more deeply, we now quantify the efficiency of excitation transport at different temperatures $T$ and site disorders. We define the efficiency of excitation transfer $\epsilon_{\tau}$ in terms of the maximum probability for the excitation to reach to the output site within a time $\tau$, following e.g.~\rref{scholak2011efficient}:
\begin{eqnarray}
\label{efficiency}
\epsilon_{\tau} = \max_{t \in [0, \tau]} |\braket{m=\text{out}}{\psi(t)}|^2.
\end{eqnarray}
The input site is \#1 as in \sref{Localization}, and there is no initial dislocation $a=X_0$, besides disorder all molecules have the same mean distance from another.

For a given choice of molecular interaction potentials and temperature, we then calculate the mean efficiency by averaging over many different kinematic configurations with the initial positions and velocities of the molecules drawn from the Maxwell Boltzmann distribution as described in \sref{AET_long}. We first calculate the maximum \bref{efficiency} for each realisation and then average these maxima over the trajectories. We finally compare two different efficiencies: $\epsilon_{\tau}^{(motion)}$, the transport efficiency in the case of mobile molecules and 
$\epsilon_{\tau}^{(static)}$, the transport efficiency in the case of immobile molecules. Note, that the latter case will still include position fluctuations according to a thermal distribution, only forces and velocities are set to zero.

\begin{figure}[htb]
\includegraphics[width=0.99\columnwidth]{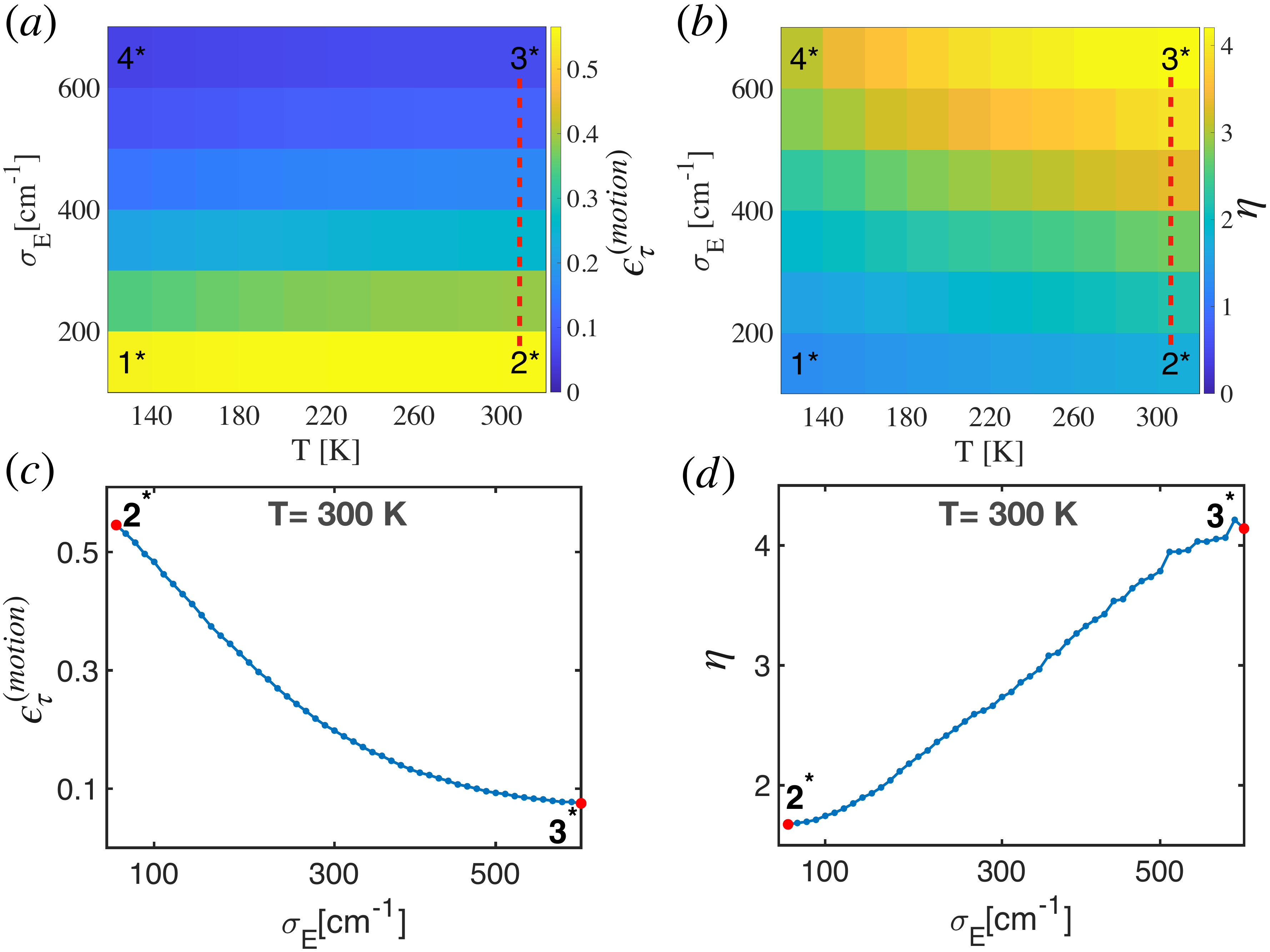}
\caption{\label{temp_longit}  (a) Mean transport efficiency $\epsilon_T^{(motion)}$ averaged over 5000 trajectories, see \eref{efficiency}, in the case of mobile molecules using a potential width according to $\alpha= 0.5 $\AA$^{-1}$. Each of the $(6\times 10)$ tiles represents a simulation result for on-site disorder strength and temperatures indicated on the axes. Labels in the corners mark cases for which we show exemplary single trajectories in \aref{Long_single_traj}, \fref{paramscan_longit_temp}. Large values of $\epsilon_T$ indicate a large probability to reach the output site from the input site. (b) Relative efficiency $\eta$ as the efficiency for the mobile scenario divided by the immobile one, see \eref{relative_efficiency}. In cases with $\eta>1$, the motion enhances transport, more so for higher values of $\eta$. Panels (c) and (d) show higher
resolution 1D cuts along $\sigma_E$ indicated by the red lines in (a) and (b).} 
\end{figure}

\begin{figure}[htb]
\includegraphics[width=0.99\columnwidth]{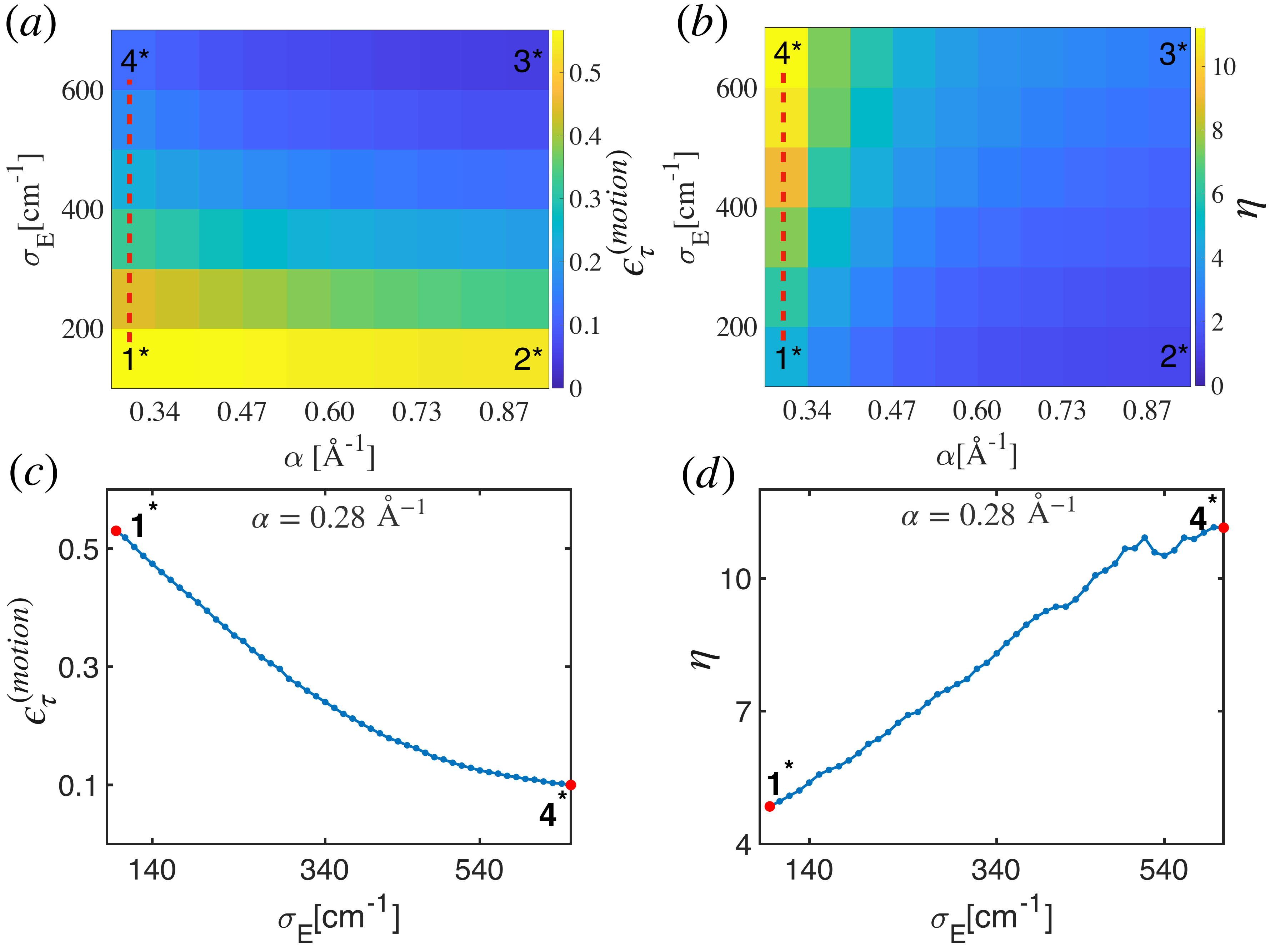}
\caption{\label{alpha_longit} The same as for \fref{temp_longit}, but we vary the width of the inter-molecular binding potential well instead of the temperature, which is fixed at $T=300 K$. Small $\alpha$ correspond to a larger width. Single trajectories for the labels in the corner of (a) are shown in \aref{Long_single_traj}, \fref{paramscan_longit_alpha}. Panels (c) and (d) show high resolution 1D cuts along $\sigma_E$ indicated by the red lines in (a) and (b). All results are averaged over $5000$ trajectories.} 
\end{figure}

The efficiency $\epsilon_{\tau}^{(motion)}$ is shown in \frefp{temp_longit}{a}, after averaging over 5000 random configurations for seven sites, for the choice $\tau=50$ ps and output site \#7. We consider a range of disorder strengths for which the aggregate transitions from de-localized excitons to strongly localized excitons for our choice of other parameters. We see that the effect of temperature on efficiency is small, within a reasonable range of temperature. This is because for the chosen $\alpha$, the accessible range of inter-molecular separations does not  significantly vary with temperature due to the tight potential, see \fref{Long_oneD} (b). To assess the impact of motion on transport for any given case, we finally resort to the relative efficiency, defined as the ratio 
\begin{eqnarray}
\label{relative_efficiency}
\eta = \frac{\epsilon_{\tau}^{(motion)}}{\epsilon_{\tau}^{(static)}}.
\end{eqnarray}
For the same scenario, the relative efficiency $\eta$ is shown in \frefp{temp_longit}{b}. In the chosen parameter range, we find relative efficiencies of up to about four, which indicates an enhanced transport in the mobile system compared to the static one.
Nextly, we examine how this enhancement is affected by the width of the potential well that binds monomers to each other. The width is controlled by the parameter $\alpha$ in the Morse potential, \bref{Morse_potential}. For small $\alpha$, the width of the well is increased, so we expect larger excursions of inter-molecular separations than for large $\alpha$. These may overcome the localization of excitons due to the accompanying strong variation of dipole-dipole interaction strengths. The efficiency for the mobile system is shown in \frefp{alpha_longit}{a}, single trajectories for the four parameter sets in the corners can be found in \aref{Long_single_traj}, \fref{paramscan_longit_alpha}. If the well is narrow, the transport efficiency remains smaller than for the case of wider well. This suggests, that larger dynamically accessible position deviations of the molecules enhance transport. Particularly for quite soft inter-molecular binding corresponding to $\alpha=0.3$ \AA$^{-1}$ in \fref{alpha_longit}, we see a marked impact of motion on excitation transport. Note however, that commonly encountered Morse potentials more typically correspond to $\alpha=0.8$ \AA.

We have chosen the relatively small system size of 7 monomers for numerical convenience. Extending this to larger system of e.g.~13 molecules, we find that the relative efficiency $\eta$ increases with system size, for otherwise identical parameters. This is expected since the localisation length from energy disorder would remain constant, hence end-to-end transport without motion will be more strongly suppressed for larger chains. In contrast, the mode of adiabatic excitation transport as in \fref{adiabatic_long} does not worsen much for larger chains, hence the relative improvement provided by motion could be larger. We have also considered more symmetric simulations with an excitation starting in the centre of the chain and exploring when it reaches either end, but allowing for an equal transport distance these are computationally more challenging and did not produce qualitatively different results.

\subsection{Transport adiabaticity} \label{Transport_adiabaticity_long}

While the motivation for the present work and the reason for our expectation that motion might aid transport stem from the concept of adiabatic excitation transport discussed in \sref{AET_long}, the results shown so far indicate only that motion may have a beneficial effect, but not whether this is due to adiabatic processes. 
When looking at individual trajectories in more detail, which we show in \aref{Long_single_traj}, it appears that the quantum dynamics of the excitation contains both, adiabatic periods as well as non-adiabatic transitions. We however do find, that the parameter space regions with the largest relative efficiencies $\eta$, are more adiabatic than others, see \aref{allowed_jumps}. 

To quantify a positive impact of adiabatic following of an exciton state on transport more precisely, we now employ the adiabaticity measure proposed in \cite{pant:adiabaticity}. Consider 
\begin{align}\label{adiab_measure}
t_n(t)&=\sum_{k,k'}\big[ \left(\dot{\tilde{a}}_{k'}(t)e^{-i \tilde{b}_{k'}(t)} \: \tilde{c}_k(t) +  \tilde{c}^*_{k'}(t)\:\dot{\tilde{a}}_k(t)e^{i\tilde{b}_k(t)}  \right) \:d^{(k)}_n(t)\: d^{(k')*}_n(t)\CR
&+ \tilde{c}^*_{k'}(t)\: \tilde{c}_k(t) \: \left( \dot{d}^{(k)}_n(t)\: d^{(k')*}_n(t) + d^{(k)}_n(t)\: \dot{d}^{(k')*}_n(t) \right) \big],\\
\sub{T}{adiab}(t)&= \int_0^t  dt' \:\: \sub{t}{out}(t'),
\end{align}
where $d^{(m)}_n(t)=\braket{n}{\varphi_m(t)}$ is the component amplitude in basis state $\ket{n}$ for system eigenstate $\ket{\varphi_m(t)}$ and we have decomposed 
the adiabatic amplitude $\tilde{c}_k(t)$ in its polar representation $\tilde{c}_k(t)=\tilde{a}_k(t) e^{i \tilde{b}_k(t)}$, with $\tilde{a}_k,\tilde{b}_k\in \mathbb{R}$.

As discussed in \cite{pant:adiabaticity}, \bref{adiab_measure} is obtained by taking the time-derivative of the population on a specific molecule $p_n(t)=|c_n(t)|^2$ and then removing the contribution due to beating between exciton eigen-states, which would also be present in the absence of motion. What remains is a measure for the site population change due to variations of the eigenstates. By construction, we have $\sub{T}{adiab}(t)=1$ at a time $t$ when the excitation has reached the output site such that this transfer was entirely due to adiabaticity. For further demonstrations and benchmarks we refer to \cite{pant:adiabaticity}.

\begin{figure}[htb]
\includegraphics[width=0.99\columnwidth]{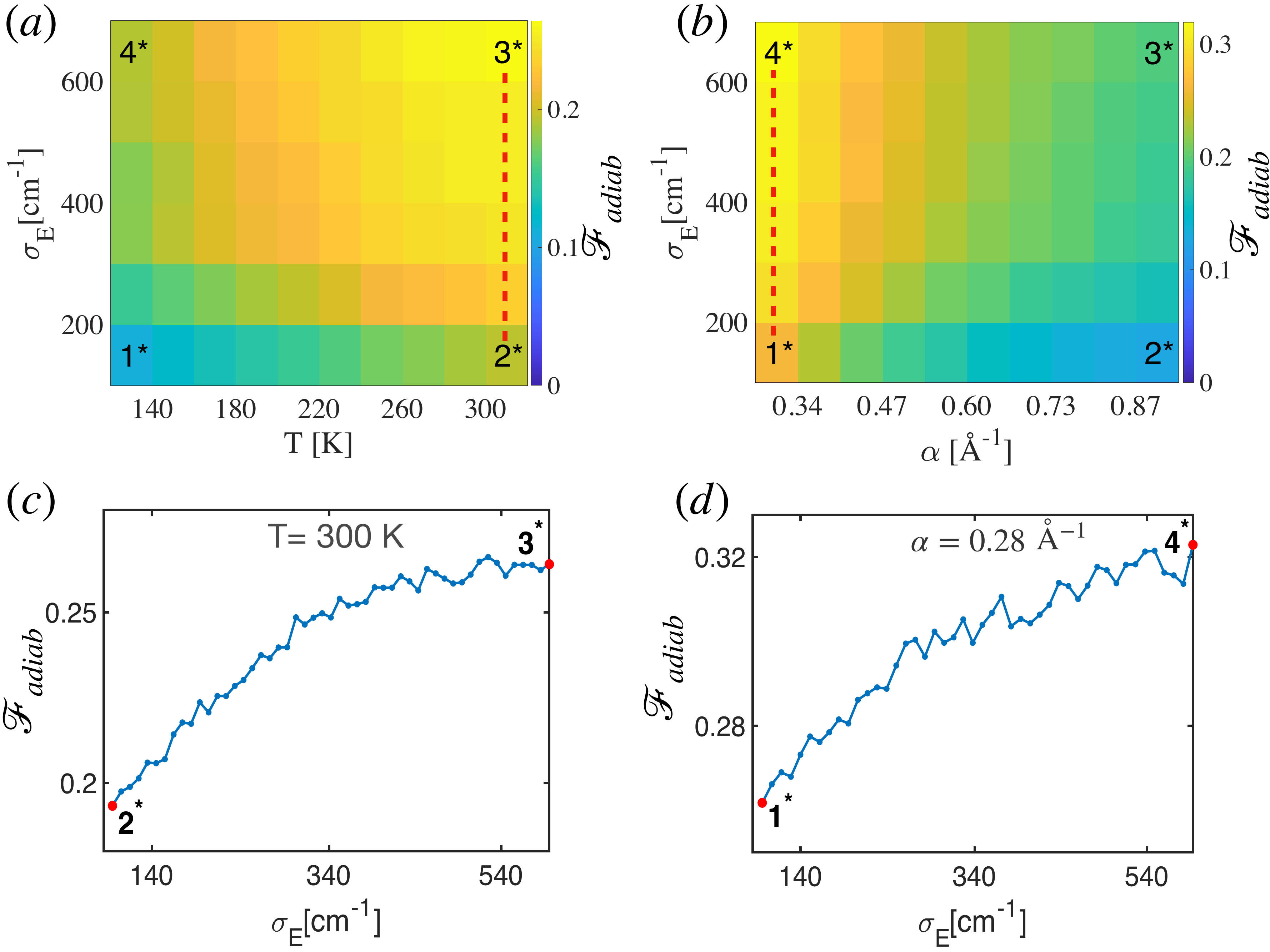}
\caption{\label{paramscan_longit_adiabaticity} Quantification of transport adiabaticity. (a) Fractional adiabaticity $\sub{\cal F}{adiab}$ (defined in the text) for the same parameters as in \fref{temp_longit} and in (b) corresponding to \fref{alpha_longit}. Panels (c) and (d) show higher resolution 1D cuts along $\sigma_E$ indicated by the red lines in (a) and (b). All results are averaged over 5000 trajectories.} 
\end{figure}
Let us define $\sub{t}{max}$ as the earliest time at which the population maximum is found on the output site, which then enters the transport efficiency evaluation \bref{efficiency}. 
With this we finally define the fractional adiabaticity $\sub{\cal F}{adiab}=\sub{T}{adiab}(\sub{t}{max})/\epsilon_\tau$. We expect $\sub{\cal F}{adiab}$ to be of order unity if the transport is fully adiabatic, and to give an indication of the relative importance of adiabatic transport otherwise. We show $\sub{\cal F}{adiab}$ in \fref{paramscan_longit_adiabaticity} for the same parameters as \fref{temp_longit} and \fref{alpha_longit}. We see that parameter space regions with larger fractional adiabaticity roughly co-incide with those showing larger relative transport efficiencies. While this suggests a possible link between the efficiency of the transport and adiabatic motion of the molecules, note that the fractional adiabaticity magnitude remains relatively small.

\section{Excitation transport by torsional motion}
\label{Rot_Cradle}

In the previous section, we have shown that while longitudinal motion of molecules along the aggregation direction can enhance the efficiency of excitation transport in the presence of disorder, this enhancement is not very large within the range of realistic parameters for molecular interaction, which are the narrowest potential explored by us. It turns out that the situation improves if the motional degree of freedom is changed from longitudinal motion to torsional motion, explored in this section. We now assume that the separation of the molecules is fixed at $3.4$ \AA, but the molecules are allowed to rotate around the aggregate $X$-axis within the $YZ$-plane. Any possible  relative tilt of the molecules out of this plane is ignored for simplicity. 

\begin{figure}[htb]
\includegraphics[width=0.99\columnwidth]{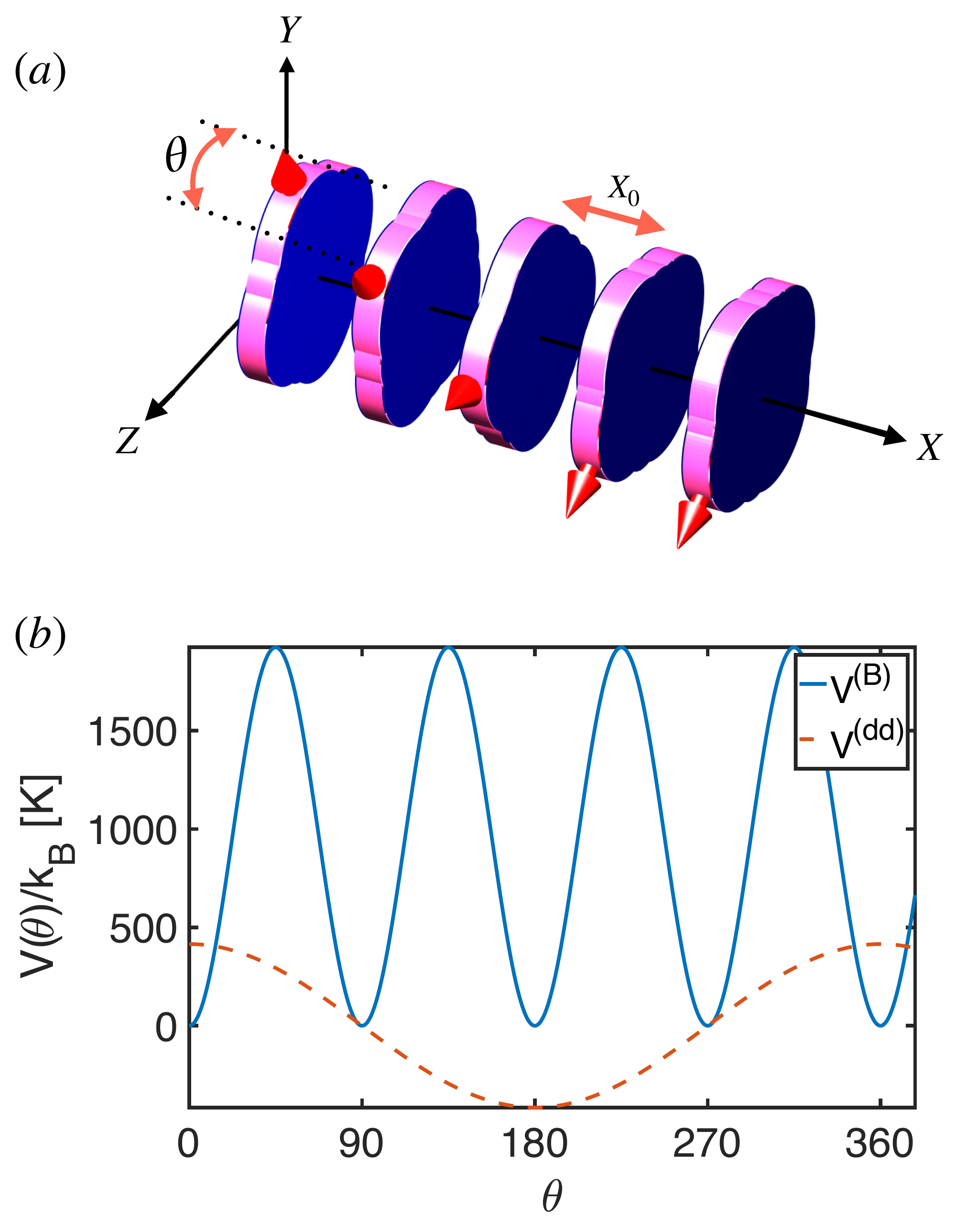}
\caption{\label{sketch_rot} (a) Sketch of the central units in a 1D self-assembled chain of dyes, which can slightly rotate about the aggregation axis.
(blue disk) molecules, (red arrow) direction of the transition dipole moment between $\ket{g}$ and $\ket{e}$.  (b) (solid blue line) Periodic potential \bref{Rot_potential} for the torsional motion of the molecules in (a), with a periodicity of $90^\circ$ due to some assumed approximate four-fold symmetry of each molecule. (red dashed) Strength of dipole-dipole interactions \bref{dip-dip_rot} for orientation.}
\end{figure}

The dipole-dipole interaction in \eref{dip-dip} with these constraints can be written as
\begin{eqnarray}\label{dip-dip_rot}
\sub V{mn}^{(dd)}(\mathbf{\theta}) =  \frac{\mu^2}{X_0^3} \cos{\theta_{mn}},
\end{eqnarray}
where $\theta_{mn}=\theta_n-\theta_m$ is the angle between the direction of the transition dipole axes of molecule $m$ and molecule $n$, shown as red arrows in \frefp{sketch_rot}{a}.  The transition dipole of magnitude $\mu$ is assumed spatially fixed in the plane of the molecule which can now rotate round the $X$-axis.

We assume that the molecules prefer to align at certain angles with their neighbors, which can be chemically engineered for example by the addition of appropriate side chains \cite{haedler2015long}. To describe these preferred orientation and torsional excursions around them, we employ a potential energy
\begin{eqnarray}\label{Rot_potential}
\sub{V}{mn}^{(B)}(\mathbf{\theta}_{mn}) = \frac{V_0}{2}[1-\cos{ (2K_{\theta} ( \theta_{mn} - \theta_{0} ) )}],
\end{eqnarray}
shown in \frefp{sketch_rot}{b}, where $V_0$ is the height of the potential barrier, $K_{\theta}$ determines the spacing of minima and hence the symmetry, and $\theta_0$ determines the equilibrium angle(s) and is fixed by the detailed shape of the molecule. For reasons that shall become clear shortly, we assume an approximate fourfold symmetry, hence $K_{\theta}=2$. The symmetry should not be perfect though, to justify the use of a single excited electronic state per molecule, and hence a well defined direction for the transition dipole moment.

In order to allow the potential \bref{Rot_potential} to control equilibrium orientations of the molecules, and not be overwhelmed by a torque from dipole-dipole interactions in \bref{dip-dip_rot}, 
we have changed the dipole strength to $\mu = 0.6$ a.u., which is about half of the value taken in \sref{Long_Cradle} for longitudinal motion.

After fixing the desired symmetry, we nextly adjust the torsional potential strength $V_0$ such that the angular spread $\Delta \theta$ in thermal equilibrium at $T=300$K is 
 $\Delta \theta = 8^{\circ}$, roughly matching angular spreads modelled in \rref{saikin2017long}. This results in the maximum value of the potential $V_0/k_B= 1923 K$. While this potential in principle still allows full rotations of the molecules, near room temperature molecules will almost exclusively perform small torsional oscillations about the minima of \bref{Rot_potential}.

We model torsional dynamics analogous to the case of longitudinal motion, except that instead of the positions $X_n$, the angles $\theta_n$ are allowed to evolve.
The classical equations of motion for the angular displacement of molecules read
\begin{eqnarray}
\label{newton_rotat}
I \frac{\partial^2}{\partial t^2}{\theta}_m = - \nabla_{{\theta}_m} U_s({\theta})  - \sum_n \nabla_{{\theta}_m} V_{mn}^{(B)}.
\end{eqnarray}
See \aref{MOI_rot} for the our simple estimate of a representative moment of inertia $I$ for dye molecules, based on CBT. As before, the exciton dynamics is obtained by expanding the total wavefunction in diabatic states 
\begin{eqnarray}\label{EoM_exciton_rot}
i \frac{\partial}{\partial t} {c}_m = \sum_{n=1}^{N} H_{mn}[\mathbf{\theta}_{mn}(t)] {c}_n,
\end{eqnarray}
where $H_{mn}[\mathbf{\theta}_{mn}(t)]$ is the matrix element for the electronic coupling in \eref{single_exciton_Hamiltonian}, with dipole-dipole interaction given by \eref{dip-dip_rot}.
 
\subsection{Idealized adiabatic excitation transport} \label{AET_rot}

As in \sref{AET_long}, we begin with the basic demonstration that molecular torsion can cause transport in principle.
We assume that the orientations of all molecules are at the equilibrium of the torsional potential \eref{Rot_potential}, with angle between adjacent axes of $\theta_0 = 70^{\circ}$. Now if we decrease the angular separation between the direction of the dipole moments of first two molecules the dipole-dipole interaction between these two molecules is stronger and the excitation will again get localised on them as in \bref{adiab_estate}. The angular distribution and the angular velocity distribution of each monomer is again a Gaussian with standard deviation $\sigma_{\theta}$ and $\sigma_{\omega}$ respectively, disorder is set to zero, $E_m = 0$. Similar to \sref{AET_long}, the standard deviation in angular position ($\sigma_{\theta} = 8^{\circ}$) is small compared to the angular offset between the first two molecules ($\Delta{\theta} = 30^{\circ} $). The dynamics of excitation transport is now obtained by solving \eref{newton_rotat} and \eref{EoM_exciton_rot} as a coupled system. 

\begin{figure}[htb]
\includegraphics[width=0.99\columnwidth]{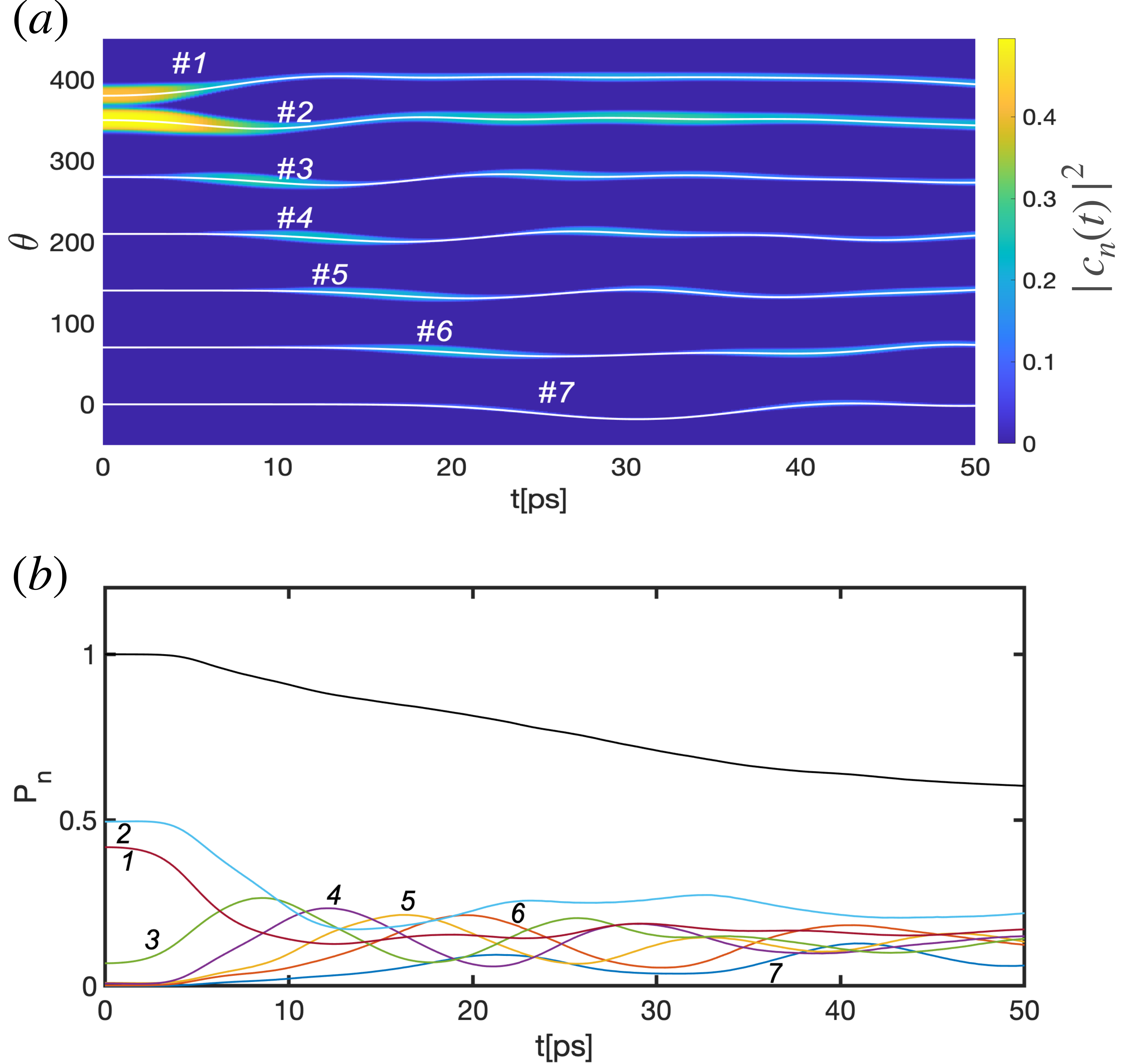}
\caption{\label{adiabatic_rot} Exemplary adiabatic excitation transport by molecular rotations. (a) (solid white) Mean orientation angles of individual molecules, with site index $n$ as indicated. (color shading) Mean electronic excitation probabilities or diabatic populations $|c_n|^2$ are indicated by color around the respective trajectory. (b) (colored numbered lines)  Individual diabatic populations $|c_n|^2$, (black solid) adiabatic population on the initial surface. Results are averaged over $10^5$ trajectories.}
\end{figure}

The resultant excitation transport is show in \fref{adiabatic_rot}, averaged over $10^5$ trajectories. The dislocation at the end will cause a repulsive torque on molecule 2 which causes it to rotate its axis towards that of molecule 3, increasing the dipole-dipole interaction between those two and carrying the excitation with it, as in \sref{AET_long}. Again the scheme proceeds over several sites.
Compared to the scenario of \sref{AET_long}, we see a larger reduction of adiabatic population as the black line in \frefp{adiabatic_rot}{b}, with a corresponding larger drop of the fidelity of transport.
%
\subsection{Transport efficiency} \label{RT}
%
In this section we explore to what extent \emph{torsional} motion of the molecules during energy transport can overcome disorder. We again take a chain of seven molecules, however now assume orthogonal dipole moment axes of adjacent molecules, i.e. $\theta_0=\pi/2$ in \frefp{sketch_rot}{a}. We chose this angle since dipole-dipole interactions at the precise equilibrium position vanish, which will necessarily enhance the \emph{relative} impact of random molecular rotations near the equilibrium orientation on dipole-dipole interactions and hence exciton states.

\begin{figure}[htb]
\includegraphics[width=0.99\columnwidth]{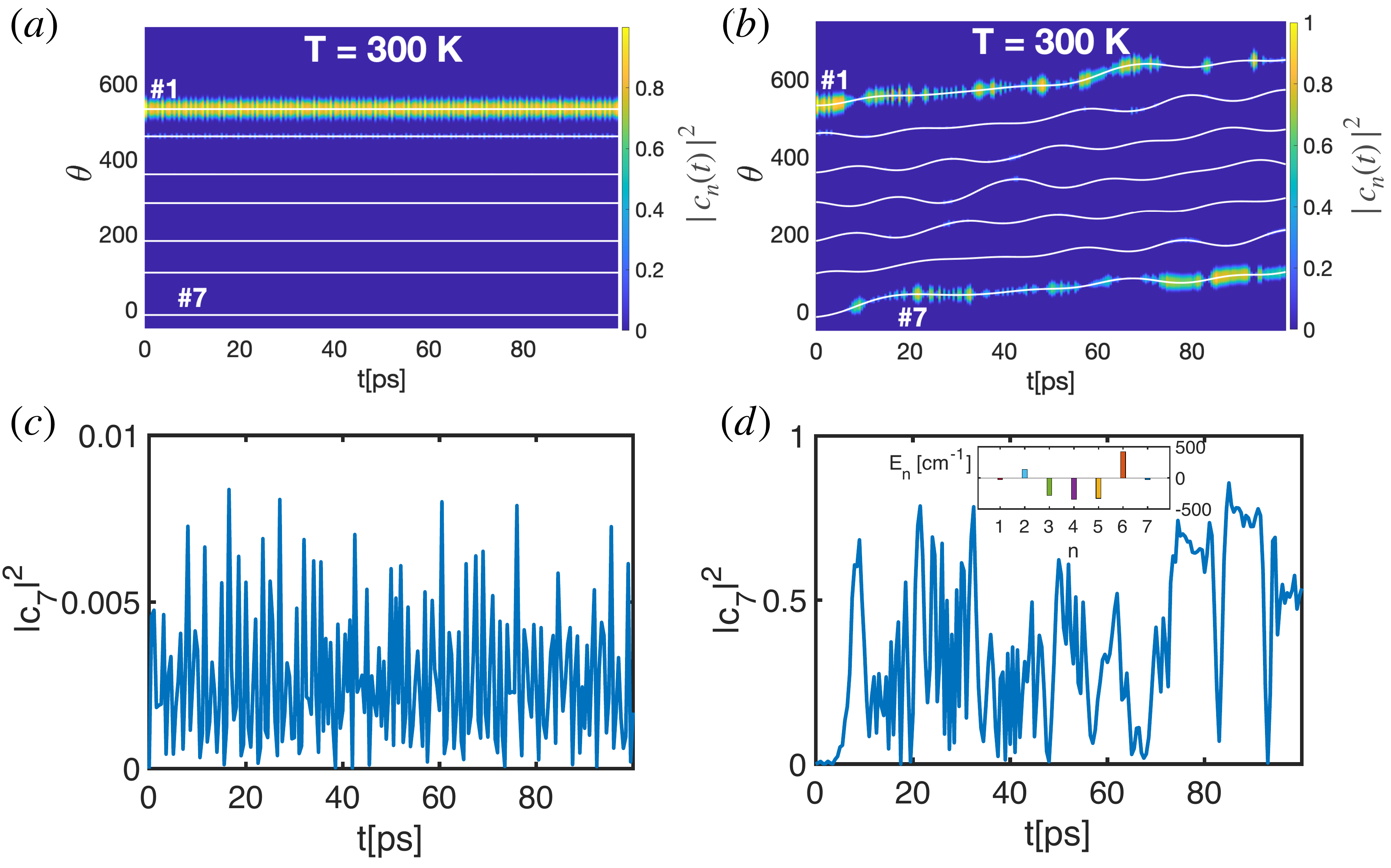}
\caption{\label{ST_rot} Excitation transport through torsional motion can overcome disorder. (a) Monomer orientations and excitation probability in the immobile case in the same style as for \fref{adiabatic_long}. (b) The same for mobile molecules. (c) Population reaching the output site \#7 in the immobile case. (d) The same for the mobile case. The inset in (d) shows the selected single realisation of on-site energy disorder.}
\end{figure}

\begin{figure}[htb]
\includegraphics[width=0.99\columnwidth]{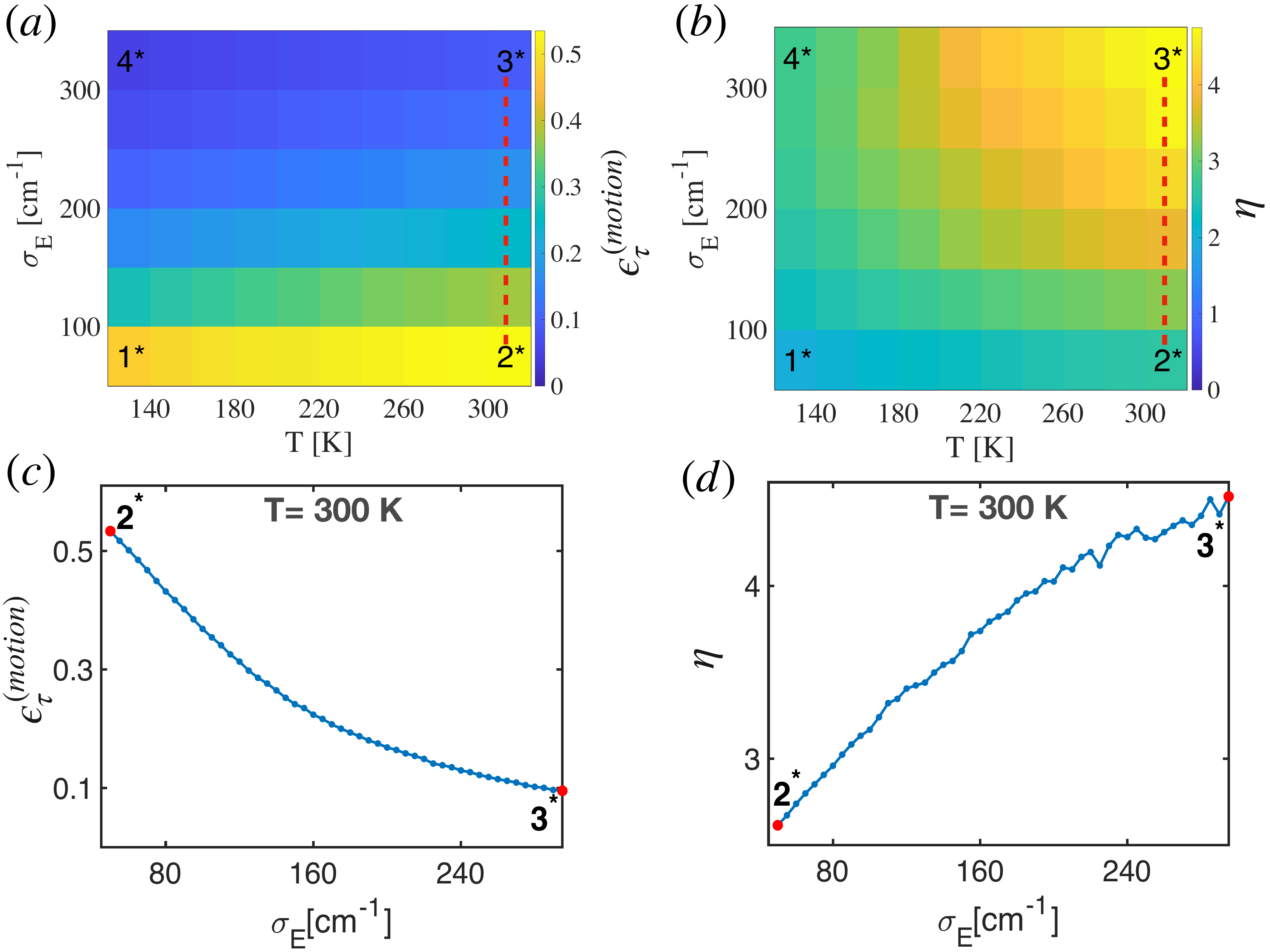}
\caption{\label{Efficiency_rot_temp} Variation of transport efficiency with disorder strength and temperature for torsional motion of molecules, for a well width of $\sigma_{\theta}= 8^{\circ}$. (a) Absolute efficiency $\epsilon_T^{(motion)}$ in the mobile case. Labels in the corners mark cases for which we show exemplary single trajectories in \aref{rot_single_traj}, \fref{paramscan_rot_temp}. (b) Relative efficiency $\eta$ as the efficiency for the mobile scenario divided by the immobile one. Panels (c) and (d) show higher resolution 1D cuts along $\sigma_E$ indicated by the red lines in (a) and (b). The results are averaged over $5000$ trajectories.} 
\end{figure}

\begin{figure}[htb]
\includegraphics[width=0.99\columnwidth]{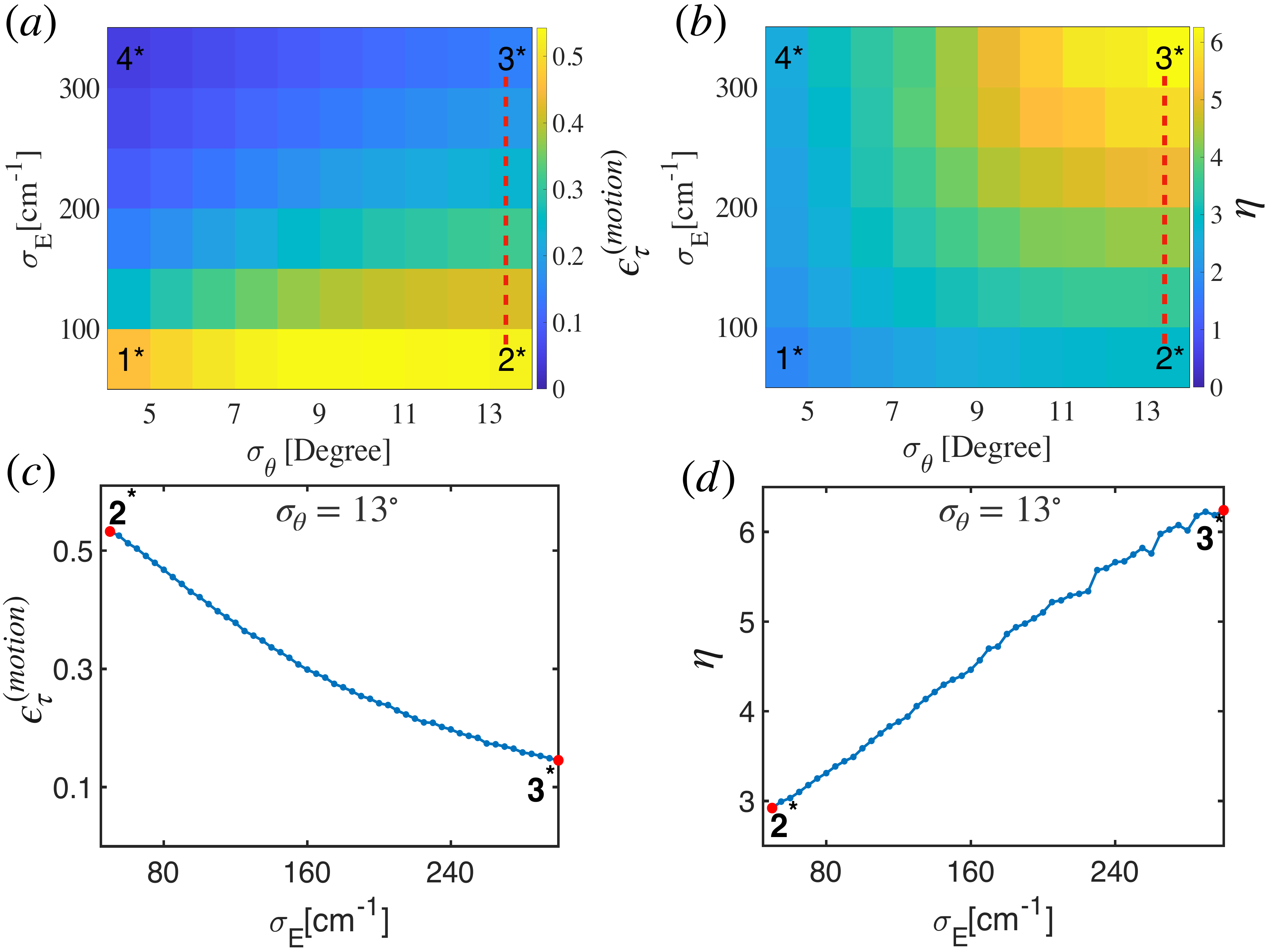}
\caption{\label{Efficiency_rot_sTheta} Variation of transport efficiency with disorder strength and well width for torsional motion of molecules at $T=300$ K. (a) Absolute efficiency $\epsilon_T^{(motion)}$ in the mobile case. Labels in the corners mark cases for which we show exemplary single trajectories in \aref{rot_single_traj}, \fref{paramscan_rot_sTheta}. (b) Relative efficiency $\eta$ as the efficiency for the mobile scenario divided by the immobile one. Panels (c) and (d) show higher resolution 1D cuts along $\sigma_E$ indicated by the red lines in (a) and (b). The results are averaged over $5000$ trajectories.} 
\end{figure}

Fluctuations of on-site-energies are again described by \eref{Energy_Distribution}. The transport of excitation for a single trajectory at room temperature for both, the mobile and the static system is shown in figure \fref{ST_rot}. As in \sref{Localization} the initial state of excitation at time $t=0$ is $\ket{\Psi(0)}=\ket{1}$. The standard deviation for orientation angles at room temperature is taken to be $\sigma_{\theta} = 8^\circ$. We see as in \sref{Localization}, that in a case where for the static system the excitation is almost completely localized on the input site, the mobile system manages to transfer 80\% of the excitation energy to the output site. Therefore also torsional motion of the molecules can help to combat localization and guide the transfer of excitation along the chain.
The effect of site disorder and temperature on efficiencies and relative efficiencies is shown in \fref{Efficiency_rot_temp}, using the same definitions as in \sref{Transport_Efficiency}. 
Here the potential \bref{Rot_potential} is fixed and the efficiency is obtained by averaging over 5000 random configurations of seven sites. We have re-adjusted the range of disorder strengths to cover the regime from de-localized excitons to strongly localized excitons for the different setting here.
In contrast to results in \sref{Transport_Efficiency}, there is now a small increase in efficiency with temperature for the mobile system. We see an even more significant increase in the relative transport efficiency \fref{Efficiency_rot_temp}{b}, compared to the case of longitudinal motion.
Nextly, we fix the temperature at $T=300$ K and vary the width of the potential well by changing  $V_0$ in the potential \bref{Rot_potential} and hence
$\sigma_{\theta}$. The results are shown in \fref{Efficiency_rot_sTheta}. 
As in the earlier section on longitudinal motion, we recover the scheme that wider potentials allowing a larger range of motion give rise to larger relative efficiencies of excitation transport.

\subsection{Transport adiabaticity} \label{Transport_adiabaticity_rot}

An inspection of the adiabaticity of individual trajectories, shown in \aref{rot_single_traj}, again shows a mixture of adiabatic and non-adiabatic contributions to the dynamics, as we had seen for the scenario with longitudinal motion along the aggregate axis. We also again find the general trend that regions in parameters space with high relative efficiency co-incide with those showing more adiabaticity, see  \aref{allowed_jumps}.

\begin{figure}[htb]
\includegraphics[width=0.99\columnwidth]{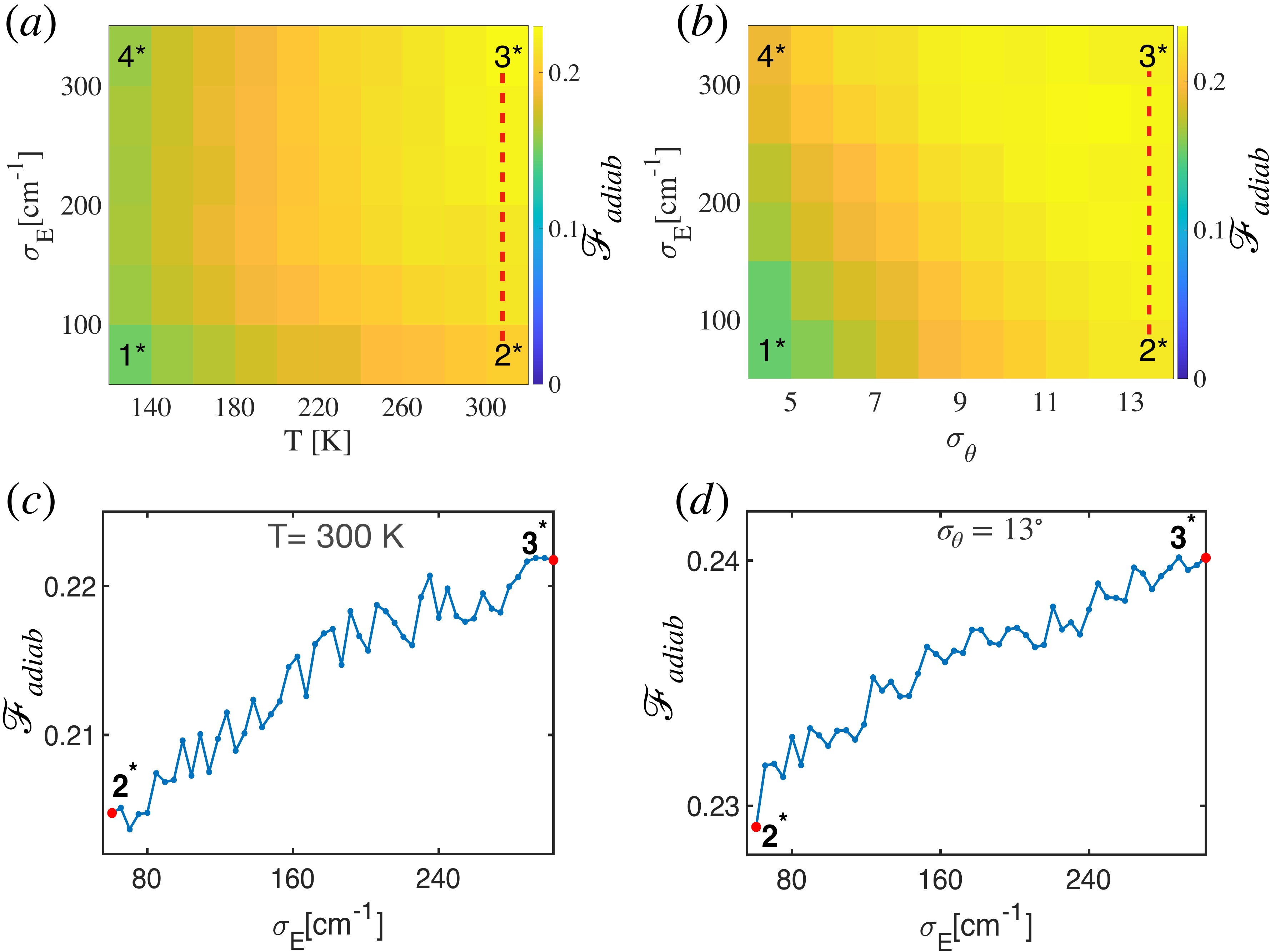}
\caption{\label{paramscan_rot_adiabaticity} Quantification of transport adiabaticity. (a) Fractional adiabaticity $\sub{\cal F}{adiab}$ (defined in \sref{Transport_adiabaticity_long}) for the same parameters as in \fref{Efficiency_rot_temp}. (b) Corresponding to \fref{paramscan_rot_sTheta}. Panels (c) and (d) show higher resolution 1D cuts along $\sigma_E$ indicated by the red lines in (a) and (b). The results are averaged over 5000 trajectories.}
\end{figure}

The same quantification $\sub{\cal F}{adiab}$ of the relevance of adiabaticity for the enhancement of quantum transport through motion that we discussed in \sref{Transport_adiabaticity_long} is shown in \fref{paramscan_rot_adiabaticity} for the case of torsional motion of monomers. Similar to our results for longitudinal motion, the fractional adiabaticity is large in the regime where the relative efficiency is large, suggesting a link of the increase in the efficiency of the transport and adiabatic rotation of the molecules.
%

\section{Conclusions and outlook} \label{Conclusions}
%
We have explored how mechanical motion of the molecules in an aggregate affects the efficiency of excitation transport compared to an immobile scenario. We find a motion-induced  enhancement of the excitation transfer efficiency over static configuration in the presence of on-site energy disorder for both, longitudinal motion of molecules along the aggregate axis and rotational or torsional motion of them in the plane orthogonal to that axis, in small systems of up to $N=7$ monomers. We conclude that a strong connection between the motion of the molecules and the propagation of the electronic excitation has the potential to increase the efficiency of excitation transfer significantly, in cases where the coupling to internal molecular vibrations can be neglected.

We separately simulated longitudinal motion of monomers along the aggregation axis and torsional motion in a plane orthogonal to that axis, and found that the torsional mode will have more beneficial impact on transport due to larger variations in interaction strengths that are realistically accessible. Additional simulations involving both torsional and translational degrees of freedom show qualitatively unchanged results compared to those presented here.

While a possible cause of enhancement of transport can be found in adiabatic excitation transport \cite{wuster2010newton} as demonstrated in \sref{AET_long} and \sref{AET_rot}, for most of the inspected parameter space the situation is less clear, with dynamics exhibiting both: periods where eigenstates are adiabatically followed, interspersed with sudden non-adiabatic transitions. For electronic dynamics involving superpositions of excitons and thus transport from site to site also in the absence of any motion, a clear link between transport and adiabaticity or its absence is difficult to establish. To nonetheless quantify the importance of adiabaticity for the motional enhancement of transport, we have employed the adiabaticity measure proposed in \cite{pant:adiabaticity}. We find that adiabatic contributions are significant but not likely crucial for transport enhancement, but regions of high adiabaticity co-incide with those showing efficient transport.

In this article, we have ignored the effect of \emph{intra}molecular vibrations in excitation transport, although these frequently play a crucial role in transport \cite{caruso2009highly,ritschel2015non, roden2009electronic, ritschel2011efficient, roden2012accounting,qin2014dynamics, ai2012complex, olaya2008efficiency, liang2010excitation, ghosh2011quantum}. In the next step of this exploration, we will thus extend the quantum dynamics calculation for excitation transport to an open-quantum-system technique, such as non-Markovian quantum state diffusion \cite{suess2014hierarchy, diosi1997non, diosi1998non, roden2009influence}, which will be coupled to classical time-dependent trajectory for the molecules as in this work.
Earlier research using simpler models for motion and energy disorder than the present work suggests that motion can enhance transport even in the presence of decohering environments \cite{Asadian_2010,o2014non}.

Another significant extension to bridge the gap between these model calculations and realistic molecular systems, would be to treat energy transport beyond the dipole-dipole approximation.
For the relevant case where the intermolecular distance is comparable to the size of the molecules, higher multipole transitions play a significant role\cite{krueger1998calculation, nasiri2018resonance, craig1998molecular}. Short range excitonic couplings may significantly deviate from the dipolar form showing an exponential distance dependence \cite{Madjet_shortrangecoupl}. Since adiabatic transport discussed here relies on significant changes of the excitonic Hamiltonian with molecular positions, it might be enhanced by such effects.

For a final confirmation for realistic and technologically relevant settings, such as dye-sensitised light harvesting technology \cite{zhang2017dye, ghosh1978merocyanine}, we can replace the simple classical point particle motion of the present article \bref{newton_longit} by full fledged molecular dynamics simulations evolving all the nucleii, and the evolution of electronic states by the simple matrix model \bref{TDSE_diabatic_basis} used here with time-dependent density functional theory of the many electron aggregate system in what is known as QM-MM schemes. Such studies could then confirm whether adiabatic or motion-induced enhanced transport is an effective countermeasure to exciton trapping. Alternatively even the full quantum dynamics of the system might be tractable  using multiconfigurational techniques such as the multi-layer multiconfiguration time-dependent Hartree (ML-MCTDH) method \cite{binder2019first}. 

Finally, since this work was inspired by ideas that have originated in a quantum simulation context with cold atoms \cite{wuster2010newton}, also other features discovered in that context might be portable to a molecular setting. One prominent one is the use of conical intersections \cite{yarkony2001conical} as switches for coherence and direction of excitation transport \cite{leonhardt2014switching, leonhardt2016orthogonal, leonhardt2017exciton}.
\section*{Conflicts of interest}
There are no conflicts to declare.
\section*{Acknowledgements}
It is a pleasure to thank Alexander Eisfeld, Jan-Michael Rost and Varadharajan Srinivasan for helpful comments.
We also thank the Max-Planck society for financial support under the MPG-IISER partner group program. The support and the resources provided by Centre for Development of Advanced Computing (C-DAC) and  the National Supercomputing Mission (NSM), Government of India are gratefully acknowledged.
RP is grateful to the Council of Scientific and Industrial Research (CSIR), India, for a Shyama Prasad Mukherjee (SPM) fellowship for pursuing the Ph.D (File No. SPM-07/1020(0304)/2019-EMR-I).
%

\appendix
\section{Estimate of dye-molecule moment of inertia and rotational potential} \label{MOI_rot}

For example, in the supramolecular assembly of \rref{haedler2015long} that provided some guidance for our simple model of torsional motion in \sref{Rot_Cradle},
a single monomer consist of a CBT core attached via amide linker with 4-(5-hexyl-2,$2^{\prime}$-bithiophene)naphtalimide (NIBT). Therefore the total mass of the monomer is the sum of masses of these constituents, which amounts to $M = 1009.03$ amu. We simply assume that this mass is distributed uniformly in a square disc, with side length $a = 50$ \AA. The moment of inertia of a disc with mass density $\rho$ is  
\begin{eqnarray}\label{MOI}
I = \frac{1}{6}  \rho a^4.
\end{eqnarray}
For a uniform mass distribution $\rho =M/(a^2)$ we then find a moment of inertia of $ 2.95\times10^9$ a.u. 

The maximum of the potential barrier $V_0$ is obtained by taking the Taylor expansion of \eref{Rot_potential} about $\theta_{mn} = \theta_0$, and then defining a target angular width $\Delta \theta$
using the thermal equipartition theorem 
\begin{eqnarray}\label{MaxV0}
V_0 K_{\theta} (\Delta \theta)^2 = \frac{k_B T}{2 }.
\end{eqnarray}
We assume a typical angular spread $\Delta \theta = 8^\circ$ at temperature $300 K$ similar to \rref{saikin2017long}, from which we obtain ${V_0}/{k_B} = 1923 K$.

To obtain $\sigma_{\theta}$ at a different temperature T after this initial allocation, we again use \eref{MaxV0}, $\sigma_{\theta} = \sqrt{\frac{k_B T}{2K_{\theta}V_0}}$. The distribution in angular velocity also relies on the equipartition theorem, yielding a width $\sigma_{\omega} = \sqrt{k_BT/I}$. 

\section{Single trajectories for longitudinal motion}\label{Long_single_traj}

To provide more intuitive access to the ensemble averaged results in the main text, \sref{Transport_Efficiency}, we now additionally provide some individual trajectories along the edges of the investigated parameter space.

\begin{figure}[htb]\label{Longitudinal_Cradle_Appendix_temperature}
\includegraphics[width=0.99\columnwidth]{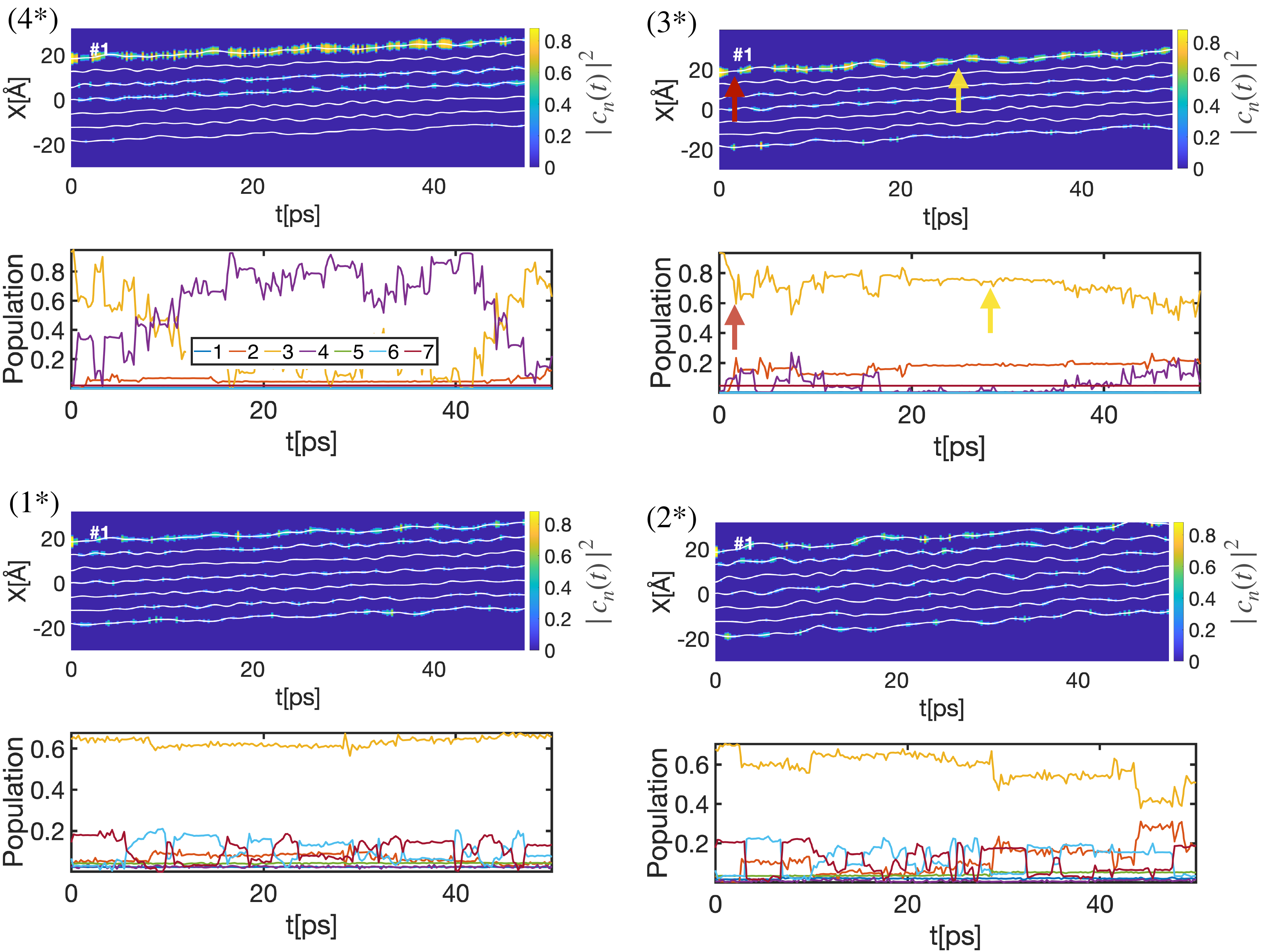}
\caption{\label{paramscan_longit_temp} Single trajectory time evolution of the position of individual molecules together with exciton dynamics for $\alpha = 0.5$ \AA$^{-1} $. The numbering refers to tags in  \fref{temp_longit} and the four different parameter sets given in \tref{tab:table1}. Top panels show molecular coordinates and diabatic populations $|c_n(t)|^2$ in the same style as \frefp{adiabatic_long}{a}. Bottom panels show the corresponding adiabatic populations $|\tilde{c}_n(t)|^2$, with legend included for case (4$^*$). Surfaces are numbered from the highest energy exciton state to the lowest. Colored arrows refer to specific dynamical events discussed in the text.
} 
\end{figure}

We see in the bottom panels (1* and 2*) of \fref{paramscan_longit_temp}, that for small energy disorder $\sigma_E$, the localization effect is weak. Thus even in the absence of molecular motion the excitation quite likely and rapidly reaches the output site, a result that no longer can be much improved upon by motion. In contrast, the upper panels (3* and 4*) with strong disorder show quite localised excitons, where for example panel \#3 then shows dynamics where this disorder has been overcome by thermal motion. 

\begin{table}[h]
\small
  \caption{Parameters for single trajectory data in \fref{paramscan_longit_temp}. Positions refer to the tags in \fref{temp_longit}}.
  \label{tab:table1} 
  \begin{tabular*}{0.48\textwidth}{@{\extracolsep{\fill}}lll}
    Positions & Parameters\\
\hline
$\#1$ & $\sigma_E = 100$ $cm^{-1}$ and $T = 120$ $K$\\
$\#2$ & $\sigma_E = 100$ $cm^{-1}$ and $T = 300$ $K$\\
$\#3$ & $\sigma_E = 600$ $cm^{-1}$ and $T = 300$ $K$\\
$\#4$ & $\sigma_E = 600$ $cm^{-1}$ and $T = 120$ $K$\\
    \hline
  \end{tabular*}
\end{table}

\begin{figure}[htb]
\includegraphics[width=0.99\columnwidth]{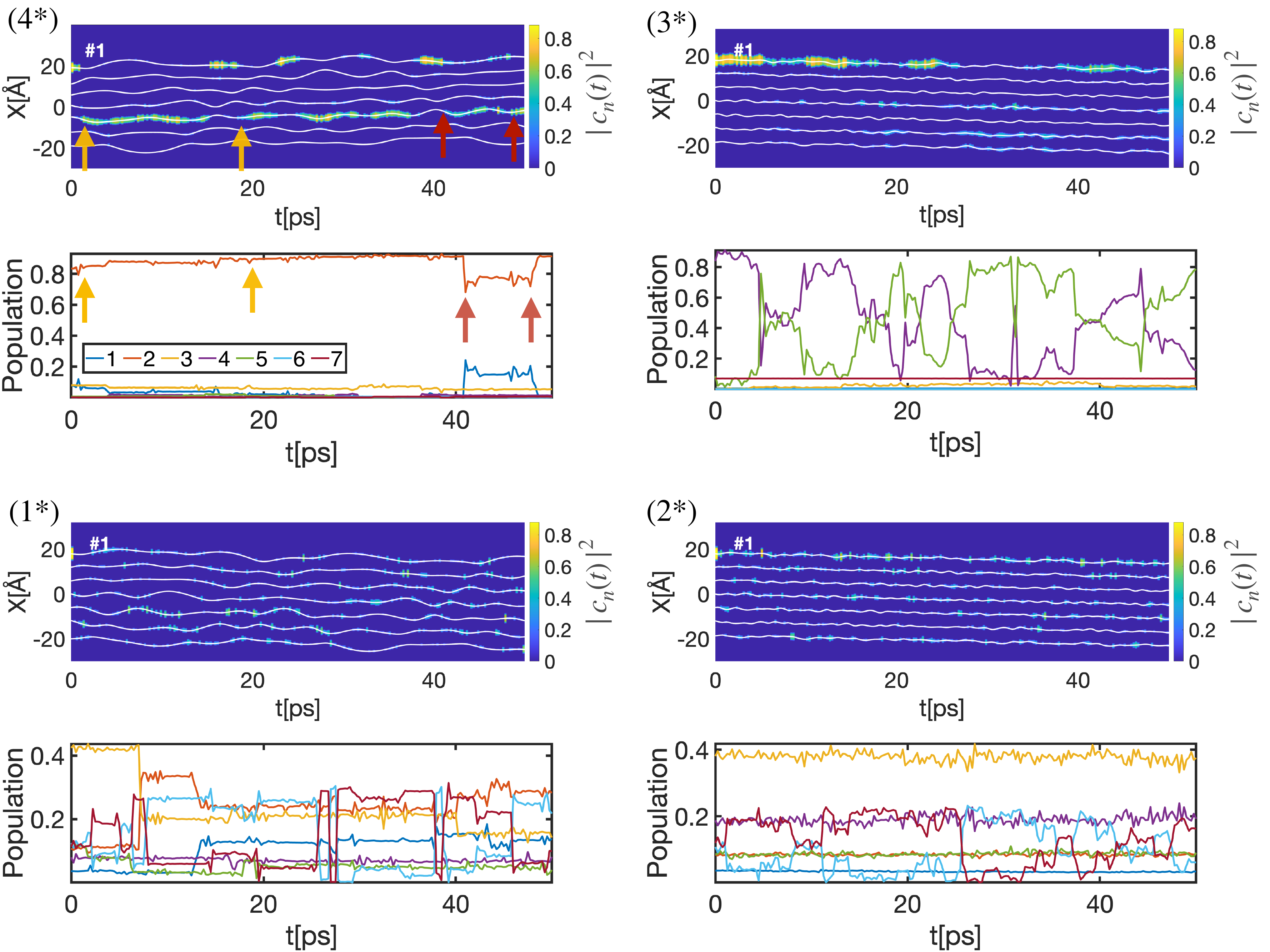}
\caption{\label{paramscan_longit_alpha}  Single trajectory time evolution of the position of individual molecules together with excitation dynamics, in the same style as \fref{paramscan_longit_temp}, but here we fix the temperature at $300 K$. The numbering refers to tags in \fref{alpha_longit} and the four different parameter sets given in \tref{tab:table2}.
Colored arrows refer to specific dynamical events discussed in the text.
}
\end{figure}

By controlling the width of the inter-molecular binding well, we can control the amplitude of excursions of the molecules. Single trajectories here show how the probability of excitation transport is influenced by changing the width as well as the site disorder. We see in \fref{paramscan_longit_alpha} that even for large site disorder the excitation energy can reach the output site with probabilities of more than $50\%$ in the mobile case. 

\begin{figure}[htb]
\includegraphics[width=0.99\columnwidth]{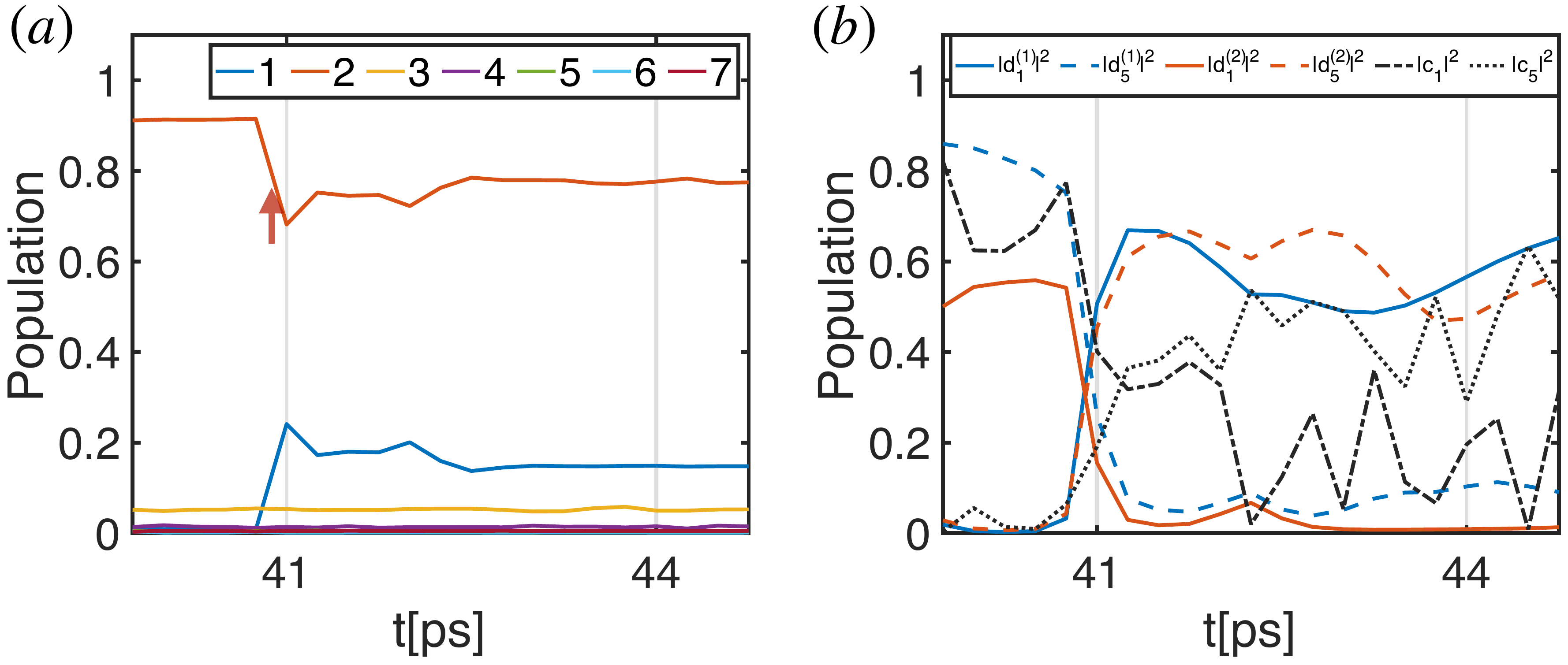}
\caption{\label{red_arrow_zoom}Only partially adiabatic transport. We show a zoom onto a red arrow in \fref{paramscan_longit_alpha} panel (4*) between $t=40$ ps and $t=44$ ps. (a) Adiabatic populations (b) The two relevant site-populations $d^{(m)}_n(t)=\braket{n}{\varphi_m(t)}$ for the two dominant excitons, see legend. We also show the total population on sites 1 and 5.
}
\end{figure}

\begin{table}[h]
\small
  \caption{Parameters for single trajectory data in \fref{paramscan_longit_alpha}. Positions refer to the tags in \fref{alpha_longit}}
  \label{tab:table2}
  \begin{tabular*}{0.48\textwidth}{@{\extracolsep{\fill}}lll}
    Positions  & Parameters\\
\hline
$\#1$ & $\sigma_E = 100$ $cm^{-1}$ and $\alpha = 0.30$ \AA$^{-1}$\\
$\#2$ & $\sigma_E = 100$ $cm^{-1}$ and $\alpha = 0.87$ \AA$^{-1}$\\
$\#3$ & $\sigma_E = 600$ $cm^{-1}$ and $\alpha = 0.87$ \AA$^{-1}$\\
$\#4$ & $\sigma_E = 600$ $cm^{-1}$ and $\alpha = 0.30$ \AA$^{-1}$\\
    \hline
  \end{tabular*}
\end{table}
Further analysis of adiabatic populations reveals that the exciton dynamics is rarely purely adiabatic but typically shows frequent non-adiabatic transitions. To assess these further, we  show adiabatic populations for the trajectories in \fref{paramscan_longit_temp} and \fref{paramscan_longit_alpha}. From these we can deduce that the evolution shows a mixture of two types of dynamics: Firstly adiabatic periods, during which adiabatic populations $|\tilde{c}_n(t)|^2$ stay nearly constant. Secondly, prominent intermittent non-adiabatic transitions, at which adiabatic populations show quite sudden significant changes.
If we manually relate either of these two features in the bottom panels with the excitation probability of each molecule shown in the top panels, we can identify two manifestations of adiabatic transport that differ in clarity: 
(i) During a period of adiabaticity, the excitation exhibits a slow transfer from one molecule to another, see \fref{paramscan_longit_alpha} panel (4*), at times indicated by yellow arrows near $t=2$ ps and $t=19$ ps. This clearly corresponds to adiabatic transport as discussed in \sref{AET_long}. (ii) Exactly coinciding with a sudden partial non-adiabatic transition, the excitation transfers from one molecule to another, see \fref{paramscan_longit_alpha} panel (4*), around $t=42$ ps and $t=48$ ps, marked with red arrows. 
Here one might suspect non-adiabatic excitation transport. A detailed inspection of the evolution of the involved eigenstates, shown in \fref{red_arrow_zoom}, however shows that transport is due to the fraction of population that remains in the same eigenstate. We thus still see adiabatic excitation transport, albeit with an efficiency reduced by the concurrent non-adiabatic transition.
%

\section{Single trajectories for torsional motion} \label{rot_single_traj}
%
Similar to the longitudinal motion scenario, we now illustrate the effect of temperature and site disorder on excitation transport with individual simulation trajectories, but fixing the molecular positions and allowing instead a rotation of the molecular transition dipole axes.

\begin{figure}[htb]
\includegraphics[width=0.99\columnwidth]{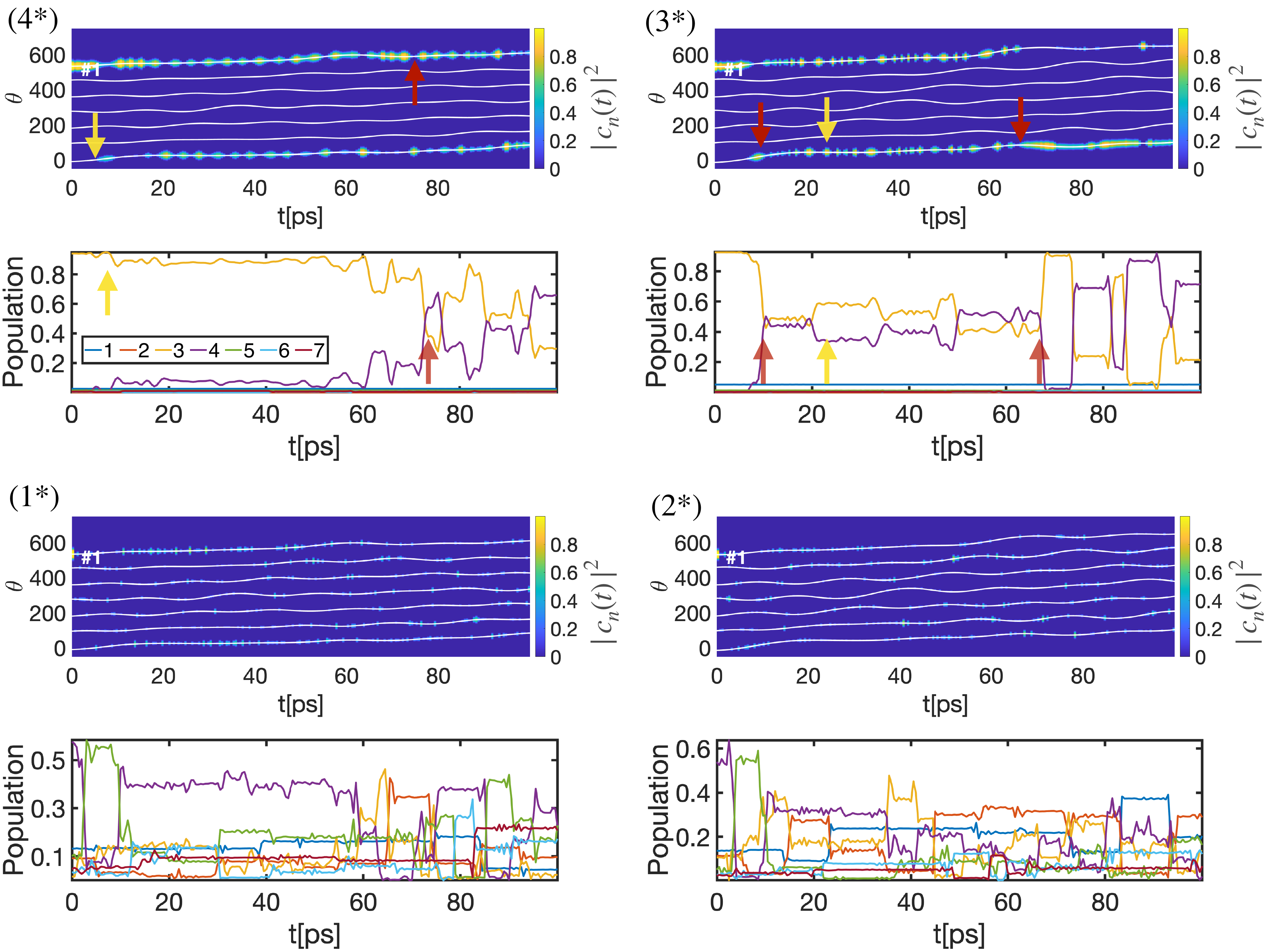}
\caption{\label{paramscan_rot_temp}  Single trajectory time evolution of the angular position and exciton dynamics for $\sigma_{\theta} = 8^{\circ}$.  The numbering refers to tags in \fref{Efficiency_rot_temp} (a) and the four different parameter sets given in \tref{tab:table3}. Even for large site disorder the excitation energy transfer to the output site is clearly seen. 
Colored arrows refer to specific dynamics events discussed in the text.}
\end{figure}

\begin{figure}[htb]
\includegraphics[width=0.99\columnwidth]{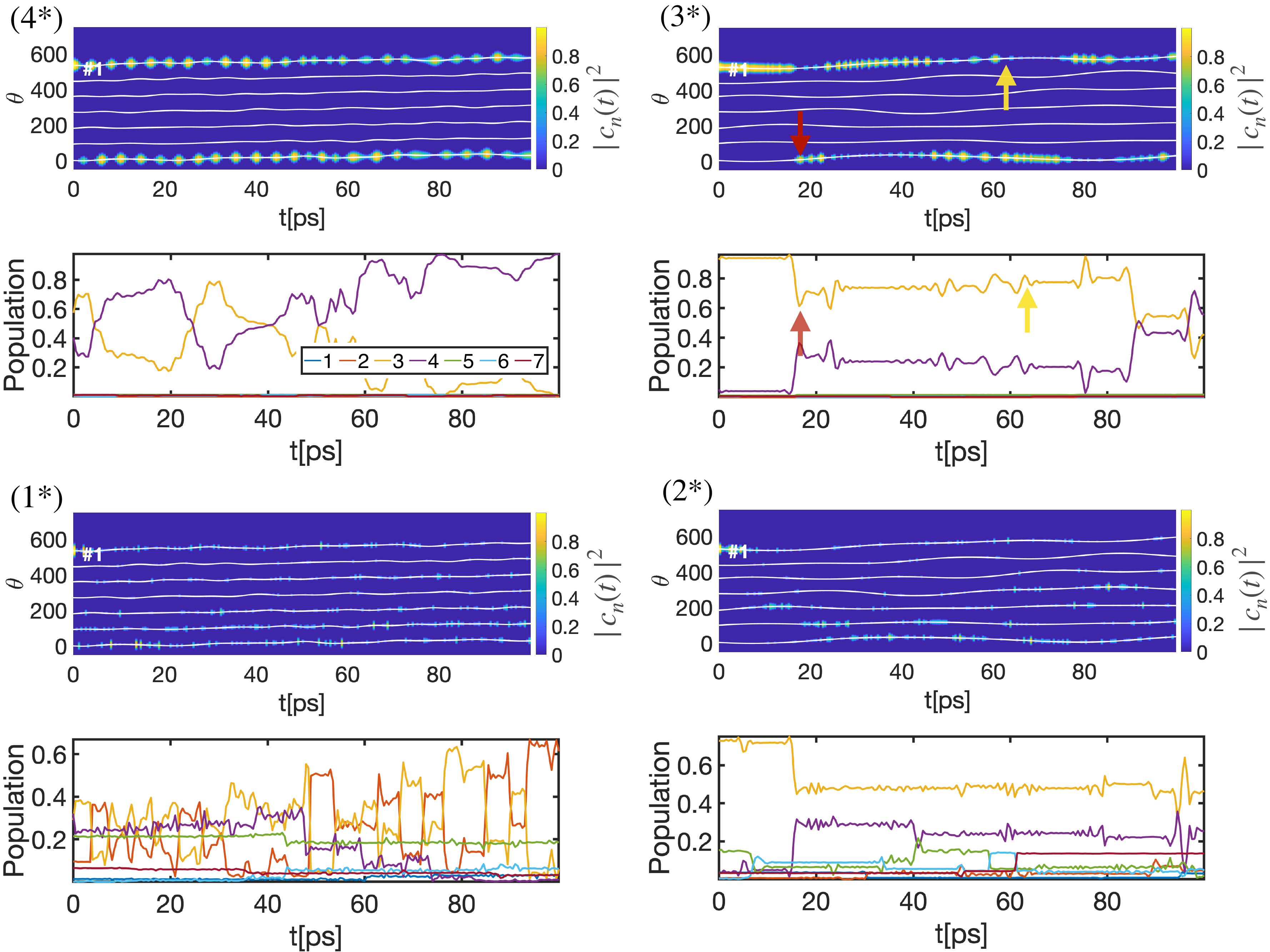}
\caption{\label{paramscan_rot_sTheta}  Single trajectory time evolution of the angular position and exciton dynamics similar to \fref{paramscan_rot_temp}, but here we fix the temperature at $300K$. The numbering refers to tags in \fref{Efficiency_rot_sTheta} (a) and the four different parameter sets given in \tref{tab:table4}. We can see a clear transport of excitation even for large site disorder. Colored arrows refer to specific dynamics events discussed in the text.}
\end{figure}

\fref{paramscan_rot_temp} shows the single trajectories at different temperatures for fixed width of the torsional potential well. When the energy disorder is small, for example in the bottom panels of \fref{paramscan_rot_temp}, the localization effect is weak and similar to the case with longitudinal motion the excitation rapidly reaches the output site with high arrival probability even in the absence of motion. In contrast, once we increase the energy disorder towards the upper panels, the excitation is more localized in the immobile system. Then, for the mobile system we see a clear transport of the exciton to the output site. 
\begin{table}[h]
\small
  \caption{Parameters for single trajectory plots in \fref{paramscan_rot_temp}. Positions refer to the tags in \fref{Efficiency_rot_temp}}
  \label{tab:table3}
  \begin{tabular*}{0.48\textwidth}{@{\extracolsep{\fill}}lll}
    Positions & Parameters\\
\hline
$\#1$ & $\sigma_E = 100$ $cm^{-1}$ and $T = 120$ $K$ \\
$\#2$ & $\sigma_E = 100$ $cm^{-1}$ and $T = 300$  $K$ \\
$\#3$ & $\sigma_E = 300$ $cm^{-1}$ and $T = 300$ $K$\\
$\#4$ & $\sigma_E = 300$ $cm^{-1}$ and $T = 120$ $K$\\
    \hline
  \end{tabular*}
\end{table}
Nextly, we provide  single trajectories for different widths of the potential in \fref{paramscan_rot_sTheta}. For wide wells (panel \#3) and high on-site disorder, when almost $90 \%$ of the exciton is localized on the first site for immobile molecules, we see that the excitation is reaching the output site with more than  $80 \%$ probability if motion is included. 
\begin{table}[h]
\small
  \caption{Parameters for single trajectory plots in \fref{paramscan_rot_sTheta}. Positions refer to the tags in \fref{Efficiency_rot_sTheta}}
  \label{tab:table4}
  \begin{tabular*}{0.48\textwidth}{@{\extracolsep{\fill}}lll}
    Positions & Parameters\\
\hline
$\#1$ & $\sigma_E = 100$ $cm^{-1}$ and  $\sigma_{\theta} = 4^{\circ}$\\
$\#2$ & $\sigma_E = 100$ $cm^{-1}$ and  $\sigma_{\theta} = 13^{\circ}$\\
$\#3$ & $\sigma_E = 300$ $cm^{-1}$ and  $\sigma_{\theta} = 13^{\circ}$\\
$\#4$ & $\sigma_E = 300$ $cm^{-1}$ and  $\sigma_{\theta} = 4^{\circ}$\\
    \hline
  \end{tabular*}
\end{table}
As in the case of longitudinal transport, we see a mixture of clear cut adiabatic transport processes and others impeded by a concurrent non-adiabatic transition. A relatively clear trajectory containing both is shown in \fref{paramscan_rot_sTheta} panel (3$^{*}$): The evolution is non-adiabatic near $20$ ps but then it is fairly adiabatic from $20$ to $85$ ps. Exactly at the moment of non-adiabaticity, the top panel shows how the excitation migrated from molecule 1 to molecule 7, due to the fraction of the population that has not changed state.  Fast oscillations in diabatic populations can be attributed to beating from a superposition of exciton states, created by the non-adiabatic transition. However note the longer time scale variations which shifts most of the excitation probability back from site 7 to site 1 around $t=25$ ps and then back again around $t=45$ ps. Neither event is accompanied by a significant change in adiabatic populations, hence we would classify these changes as adiabatic transport.

\section{Survey of adiabaticity} \label{allowed_jumps}

The relative adiabaticity measure $\sub{\cal F}{adiab}$ introduced in \sref{Transport_adiabaticity_long} provides a measure of adiabatic \emph{transport}. The less ambitious goal of quantifying the global level of adiabaticity (regardless of whether it leads to transport or not) in different regions of parameter space can be gained from our quantum-classical propagation algorithm itself and is presented in this appendix.

\begin{figure}[htb]
\includegraphics[width=0.99\columnwidth]{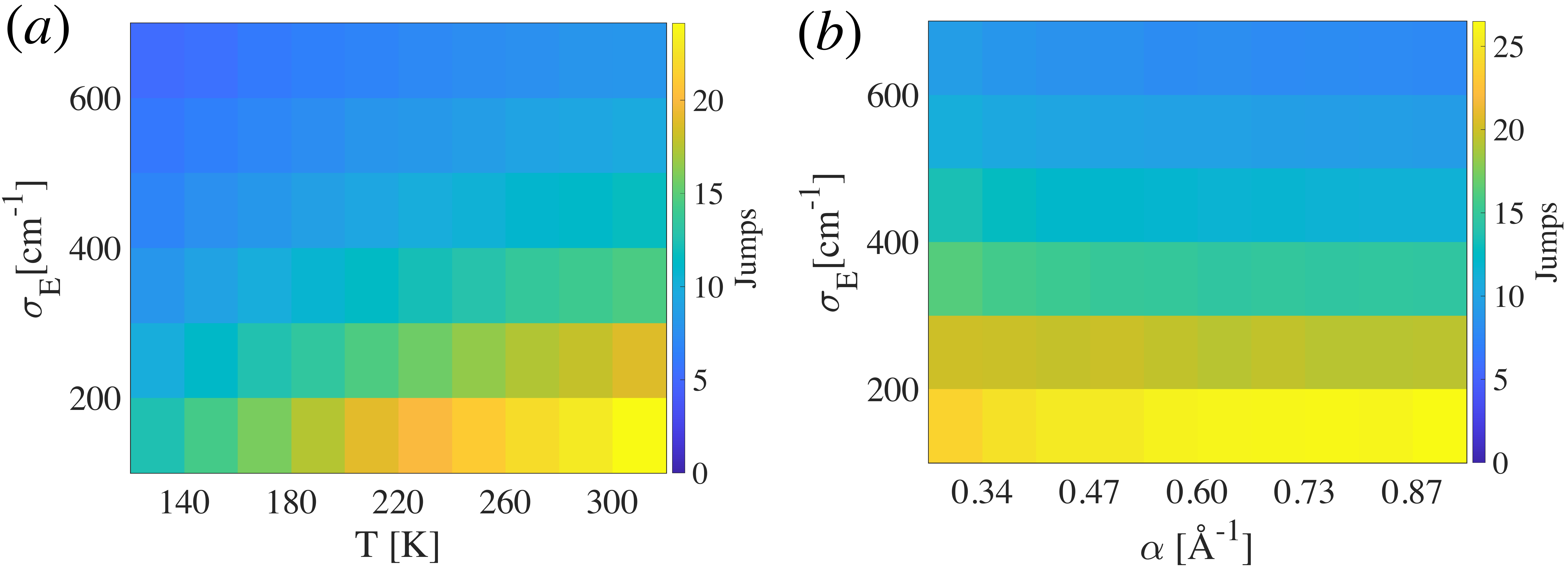}
\caption{\label{Allowed_jumps_long_vf1} Mean number of allowed surface jumps in Tully's algorithm, corresponding to non-adiabatic transitions, for longitudinal motion. (a) We vary temperatures and on-site disorder strengths. (b) We vary the width of the inter-molecular binding potential well and on-site disorder strengths. }
\end{figure}

\begin{figure}[htb]
\includegraphics[width=0.99\columnwidth]{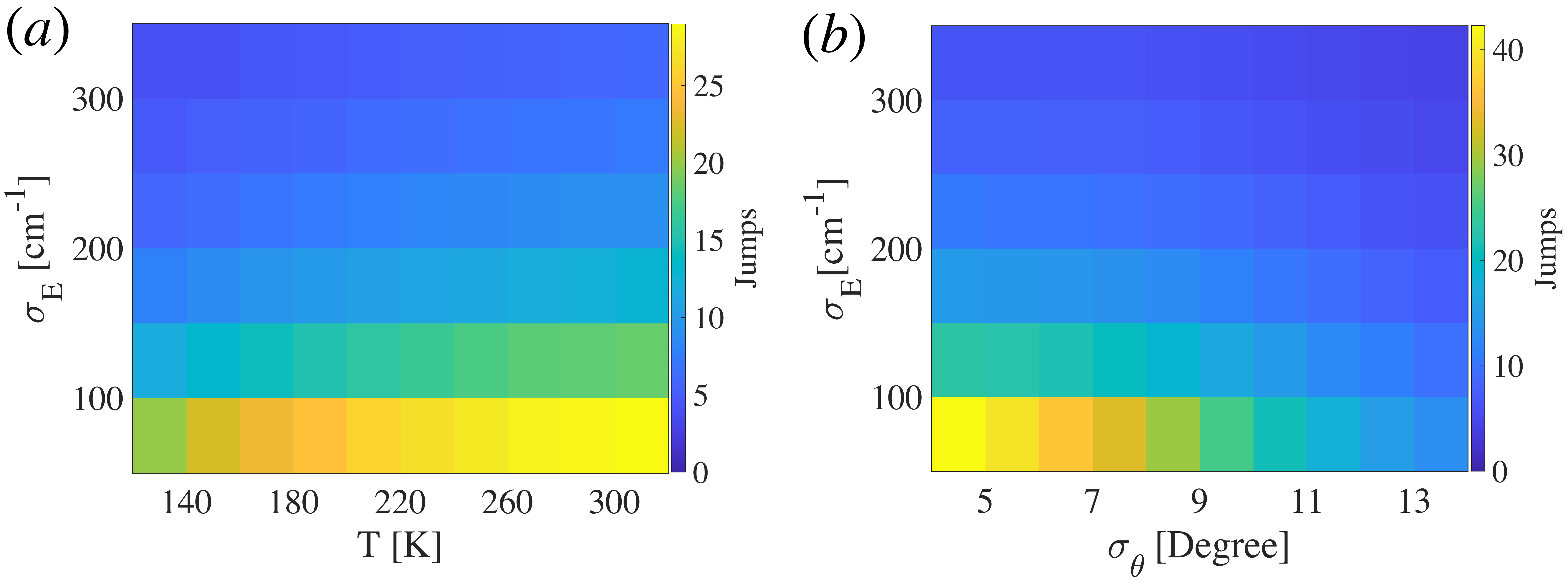}
\caption{\label{Allowed_jumps_rot_vf1} Mean number of allowed jumps similar to \fref{Allowed_jumps_long_vf1} for torsional motion. (a) For different temperatures and (b) for different widths of the well, as well as in both cases varied energy disorder strength. }
\end{figure}

The molecules move on a single adiabatic potential energy surface $m$, which may be changed via sudden jumps to another surface $n$ by non-adiabatic couplings \eref{d_mn} between surface $m$ and surface $n$. A simple estimate of non-adiabaticity is thus provided by the mean number of allowed\footnote{Jumps are not allowed if they would violate energy conservation.} jumps \cite{tully1990molecular,hammes1994proton,mobius2011adiabatic}. The mean number of jumps per trajectory is shown for longitudinal motion in \fref{Allowed_jumps_long_vf1} and for torsional motion in \fref{Allowed_jumps_rot_vf1}, for the same two cuts through parameter space discussed in the main article.
We find a larger number of jumps at high temperatures or narrow width of the potential well. This is expected since either involve faster motion of molecules, which directly increases all non-adiabatic couplings in \eref{d_mn}. When we compare \fref{Allowed_jumps_long_vf1} with \fref{temp_longit} and \fref{alpha_longit}, we find that regions of high relative efficiency coincide with those of less allowed jumps and thus more adiabatic dynamics.
Similar conclusions can be drawn from for the case of torsional motion in  \fref{Allowed_jumps_rot_vf1}. For narrow width of the potential well or high temperature, the number of allowed jumps is large due to the faster angular vibration of molecules. In contrast, there are fewer jumps when we decrease temperature or increase the width of the well or on-site disorder strength. Again, in the region of high relative efficiency in \fref{Efficiency_rot_temp} and \fref{Efficiency_rot_sTheta} the transport is more adiabatic than in other regions.
The results in this appendix point in the same direction as those in \sref{Transport_adiabaticity_long} and \sref{Transport_adiabaticity_rot}, where we show that parameter space regions where motion improves transport co-incide with higher adiabatic contributions to that transport.

\providecommand{\noopsort}[1]{}\providecommand{\singleletter}[1]{#1}%
\providecommand*{\mcitethebibliography}{\thebibliography}
\csname @ifundefined\endcsname{endmcitethebibliography}
{\let\endmcitethebibliography\endthebibliography}{}



\begin{mcitethebibliography}{79}
\bibitem[foo()]{footnote:MBdistribution}
The distribution of position and velocity of each molecule is taken Gaussian
  with the standard deviation $\sigma_x$ and $\sigma_v$ respectively . We take
  the Taylor expansion of the Morse potential about $R_0$ to get $\sigma_x =
  \sqrt{k_B T/K_e}$, where $k_B$ is Boltzmann's constant, T is temperature and
  $K_e = \Big(\partial V^{M}(\mathbf{R})/\partial R \Big)_{R=R_0} $=$2 \alpha^2
  D_e$. The distribution of velocities is also Gaussian with standard deviation
  $\sqrt{k_B T/M}$.\relax
\mciteBstWouldAddEndPunctfalse
\mciteSetBstMidEndSepPunct{\mcitedefaultmidpunct}
{}{\mcitedefaultseppunct}\relax
\EndOfBibitem
\end{mcitethebibliography}
\end{document}